\documentclass[aps,pra,superscriptaddress,twocolumn,showpacs,noeprint,notitlepage]{revtex4-2}
\usepackage{graphicx,amsfonts,times,bm,amsmath,amssymb,verbatim,ifthen,braket,xcolor,array,natbib,comment,enumerate,multirow,soul,lineno,algorithm,algpseudocode}
\usepackage[normalem]{ulem}
\usepackage[colorlinks=true,citecolor=blue,linkcolor=blue,urlcolor=blue]{hyperref}

\begin{document}

\title{Noise-resilient and resource-efficient hybrid algorithm for robust quantum gap estimation}

\author{Woo-Ram Lee}
\affiliation{Murray Associates of Utica, Utica, New York 13501, USA}

\author{Nathan M. Myers}
\affiliation{Department of Physics, Virginia Tech, Blacksburg, Virginia 24061, USA}

\author{V. W. Scarola}
\email[Email address:]{scarola@vt.edu}
\affiliation{Department of Physics, Virginia Tech, Blacksburg, Virginia 24061, USA}

\begin{abstract}
We present a hybrid quantum algorithm for estimating gaps in many-body energy spectra, supported by an analytic proof of its inherent resilience to state preparation and measurement errors, as well as mid-circuit multi-qubit depolarizing noise. Our analysis extends to a broader class of Markovian noise, employing error mitigation strategies that optimize the utilization of quantum resources. By integrating trial-state optimization and classical signal processing into the algorithm, we amplify the signal peak corresponding to the exact target gap beyond the error threshold, thereby significantly reducing gap estimate errors. The algorithm's robustness is demonstrated through noisy simulations on the Qiskit Aer simulator and demonstrations on IBM Quantum processors. These results underscore the potential to enable accurate quantum simulations on near-term noisy quantum devices without resource-intensive error correction.
\end{abstract}

\maketitle

\section{Introduction}

Quantum simulation is widely regarded as one of the most promising applications of quantum computing. The exponential scaling of Hilbert space with system size renders simulating many-body models on classical computers intractable beyond limited system sizes. Quantum-phase-estimation (QPE) algorithms~\cite{Kitaev1996, Lloyd1996, Abrams1999, Somma2002} enable quantum computers to estimate important physical quantities in quantum many-body models, such as energy eigenvalues~\cite{Somma2002, Somma2019, OBrien2019, OBrien2021, Lin2022, Wang2023, Ding2023a, Ding2023b} and gaps~\cite{Wecker2015, Zintchenko2016, Sugisaki2021, Russo2021, Matsuzaki2021}. In this context, quantum computers are believed to offer two distinct advantages over classical counterparts. Along with their well-known advantages in computational time and circuit depth, quantum computers offer a significant memory advantage by inherently encoding exponentially large Hilbert spaces within their hardware, effectively addressing the memory wall problem associated with classical random-access memory~\cite{McKee2011}.

A primary goal of many-body quantum simulation is determining low-energy eigenstates which have widespread practical values in condensed matter physics~\cite{Georgescu2014, Reiner2019, Stanisic2022}, quantum chemistry~\cite{Lanyon2010, Cao2019, McArdle2020, Nam2020, Lee2023}, and materials science~\cite{Marzari2021} due to the relevance of a ground state in determining material properties, reaction rates, and emergence of exotic quantum phases such as superconductors. In general, preparing a many-body ground state on a quantum computer is an \textit{NP}-hard problem for classical Hamiltonians~\cite{Poulin2009} and a \textit{QMA}-hard problem for general quantum Hamiltonians~\cite{Kitaev2002, Kempe2006, Aharonov2009}. Moreover, the choice of initial states significantly impacts the performance of quantum simulation algorithms and must be carefully considered when benchmarking these algorithms.  

Currently available quantum computers are noisy and limited in both qubit counts (circuit width) and gate operations (circuit depth) which pose significant challenges for conclusive demonstrations of quantum advantage in quantum simulation~\cite{Zhang2017, Arute2019, Zhong2020, Noel2022, Kim2023}. The circuit width and depth required for accurate phase estimation can be effectively reduced by employing ancilla-free variational algorithms~\cite{Santagati2018, Wang2019, Filip2024, Klymko2022}. In certain tasks, such as quantum compiling, hybrid variational algorithms have demonstrated resilience to noise, with optimal variational parameters remaining largely unaffected~\cite{Sharma2020}. These approaches, however, offer only biased solution estimates based on the variational theorem and are not inherently guaranteed to tolerate noise~\cite{Wang2024}. For unbiased (exact) simulation, certain algorithms, such as robust phase estimation~\cite{Kimmel2015} and quantum complex exponential least squares~\cite{Ding2023a, Ding2023b}, have demonstrated numerical robustness against specific types of errors~\cite{Meier2019, Russo2021, Ding2023arXiv}. Moreover, a recent study demonstrated that once a good approximation to a quantum state in the low-energy Hilbert space is identified, Hamiltonian simulation remains in the same subspace even in the presence of noise~\cite{Sahinoglu2021}.

Another promising avenue for reducing circuit width and depth is the use of hybrid quantum algorithms in conjunction with classical postprocessing~\cite{OBrien2019, Matsuzaki2021, Wan2022, Dutkiewicz2022, Lin2022}. In our prior work~\cite{Lee2024}, we introduced a hybrid quantum-gap-estimation (QGE) algorithm that integrates quantum time evolution with classical signal processing to estimate the gaps in many-body energy spectra within a tolerance. In this approach, an offline time series is constructed using a noiseless ancilla-free quantum circuit and filtered to achieve an exponential reduction in circuit depth, albeit at the cost of spectral resolution. The filter’s operating range is determined by mapping the orientations of the input qubits. This approach is highly effective for models with large spectral gaps but may face challenges in efficiently resolving small gaps in, e.g., models of spin glasses~\cite{Binder1986}. 

\begin{figure}[t]
\includegraphics[width=0.46\textwidth]{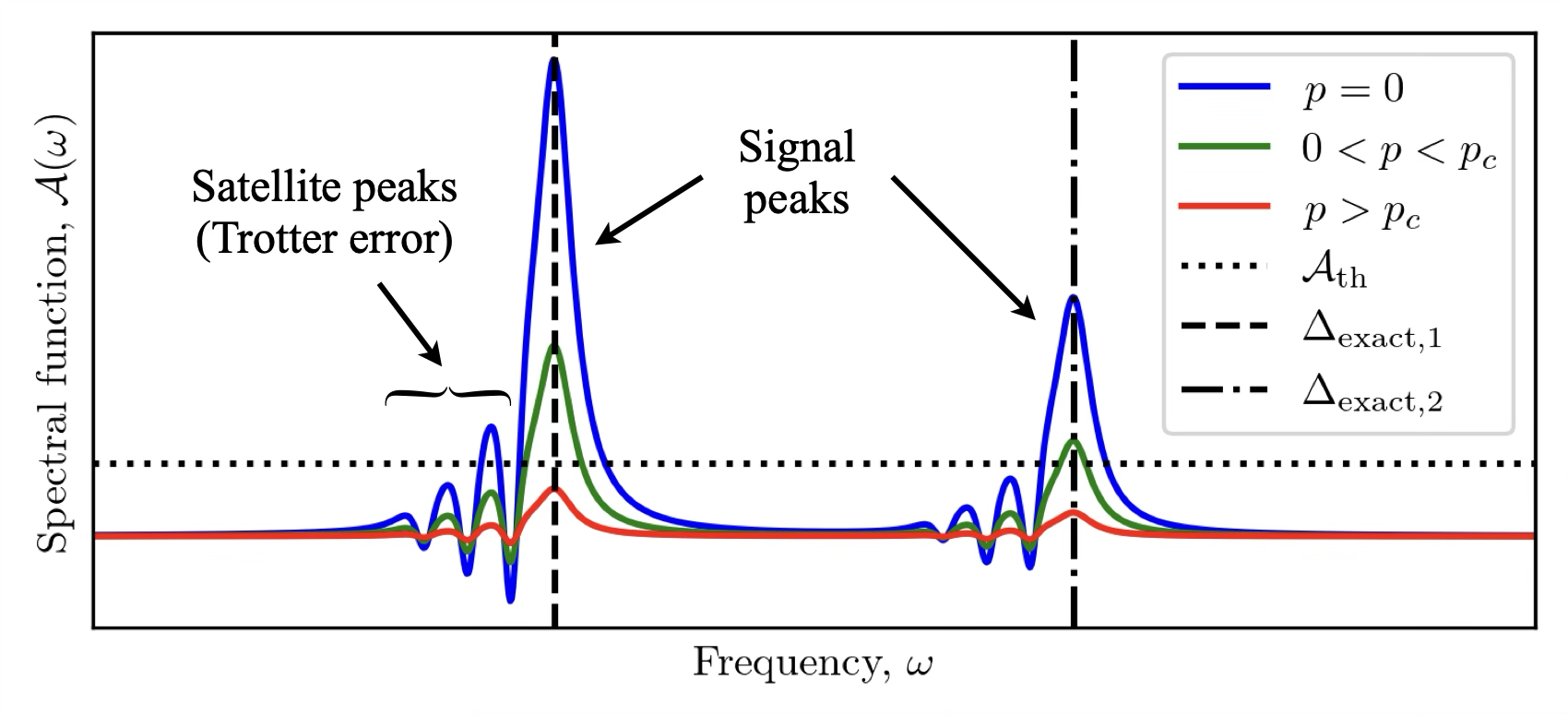}
\caption{Schematic illustrating the key features of the spectral function $\mathcal{A}$, as a function of frequency $\omega$, typically generated by the hybrid QGE algorithm. Characteristics include signal peaks (set against a flat baseline) corresponding to the exact gaps $\Delta_{{\rm exact},i}$ ($i=1,2$). When the number of Trotter steps $M$ is below the cutoff $M_c$, additional satellite peaks appear due to Trotter truncation error. The blue curve represents the noiseless result, while the green and red curves depict global suppression under circuit-level depolarizing noise with noise probability $p$ (defined as the joint probability of noise affecting individual circuits) below and above the error threshold $p_c$. The horizontal dotted line denotes the device-specific detection threshold $\mathcal{A}_{\rm th}$, influenced by factors such as the measurement shot count in the quantum process and machine precision in the postprocess.}
\label{fig_1_schematic}
\end{figure}

In this paper, we uniquely integrate trial-state optimization and classical signal processing into the hybrid QGE algorithm to enhance performance on noisy quantum computers while optimizing the utilization of quantum resources. A central aspect of this work is demonstrating that the algorithm achieves a high level of noise resilience---either inherently or through error mitigation strategies---enabling precise gap estimation for quantum many-body models. This is accomplished by iteratively updating the trial states to improve error thresholds. Figure~\ref{fig_1_schematic} illustrates a typical output from our algorithm, influenced by mid-circuit multi-qubit depolarizing noise. The output presents a spectral function with signal peaks modeled as Lorentzian shapes at the corresponding energy gaps, and smaller satellite peaks exhibiting inverse Lorentzian shapes in the presence of Trotter truncation error. Depolarizing noise uniformly suppresses both the signal and satellite peaks. The detection threshold (indicated by the horizontal dotted line) is primarily determined by the measurement shot count during the quantum process. To ensure that a signal peak remains detectable above this threshold, a sufficient number of shots must be taken, or the noise probability must be low enough to capture the true distribution, rather than being dominated by the maximally mixed state caused by depolarizing noise.

Numerical verification of our results is performed by running the algorithm on a minimal spin model using the Qiskit Aer simulator with an integrated noise model. Additionally, demonstrations on IBM Quantum processors, subject to realistic noise sources, offer further validation of our findings. This work paves the way for substantial scaling of systems, enabling unbiased quantum simulations of systems that exceed classical memory limitations on noisy intermediate-scale quantum computers~\cite{Preskill2018}. Establishing a robust foundation for demonstrating quantum advantage in quantum simulation, it sets the stage for transformative advancements in the field.

This paper is organized as follows: In Sec.~\ref{sec_hybrid_QGE}, we present an overview of the key components of the hybrid QGE algorithm, along with postprocessing techniques. In Sec.~\ref{sec_noise_resilience}, we establish the theoretical basis for the algorithm's resilience to various types of noise and explore error mitigation strategies utilizing classical signal processing. In Sec.~\ref{sec_numerical_results}, we numerically validate our findings through simulations of a minimal spin model, comparing the results across different approaches. We conclude in Sec.~\ref{sec_discussion}.

\section{Hybrid quantum-gap-estimation algorithm}
\label{sec_hybrid_QGE}

\begin{figure}[t]
\begin{center}
\includegraphics[width=0.46\textwidth]{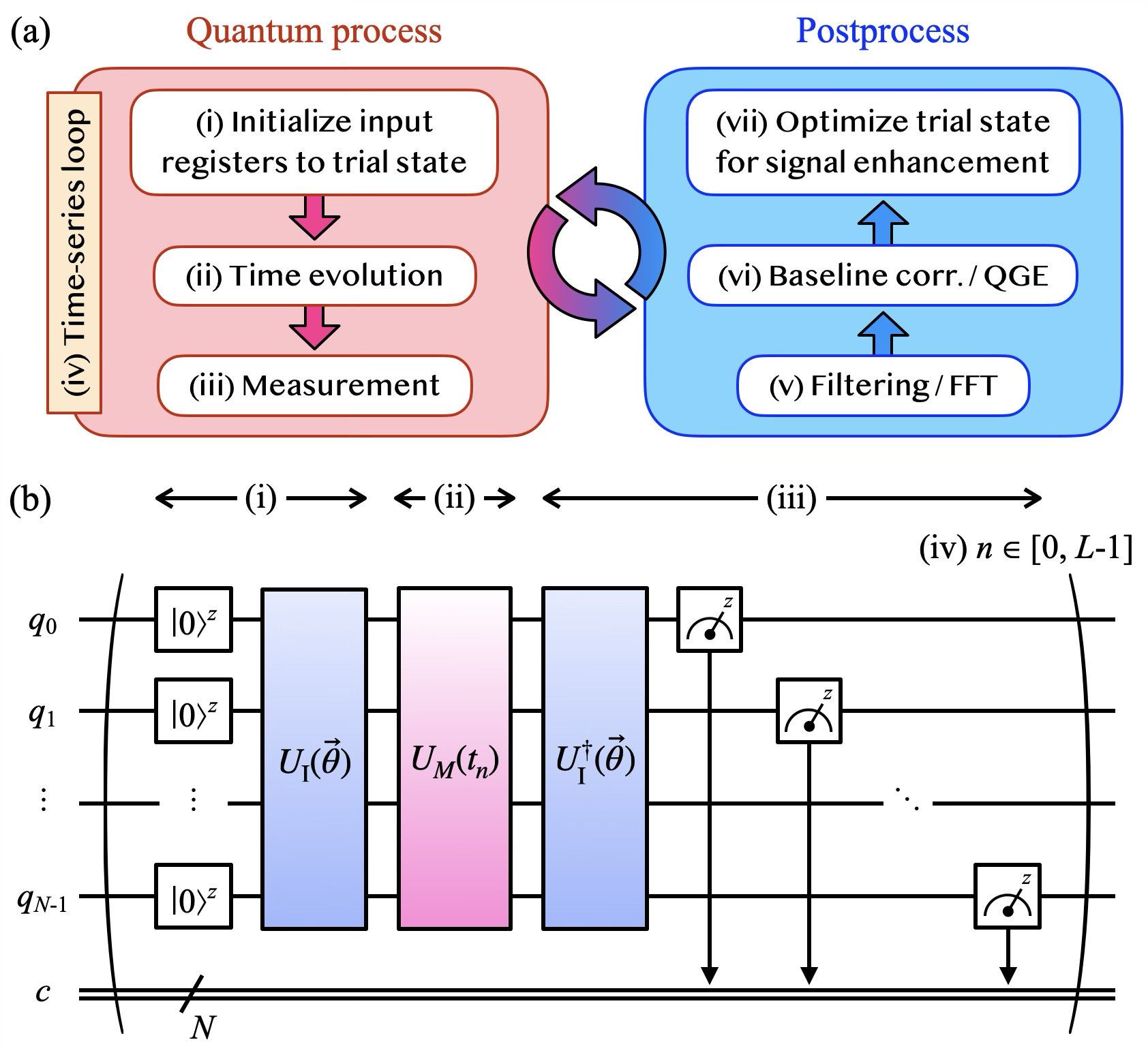}
\end{center}
\caption{(a) Flowchart for the QGE algorithm with a loop for trial-state optimization. (b) Quantum circuit for (i) initialization of input registers to a trial state, (ii) (Trotterized) time evolution, (iii) measurement. Here, unitary gates are influenced by noise. (iv) A time series of propagators is constructed from measurement outcomes and postprocessed for (v) time-series filtering followed by a FFT, (vi) gap estimation from baseline-corrected signals, (vii) trial-state optimization for enhancing signals. The loop iterates until convergence criteria are met.}
\label{fig_2_flowchart}
\end{figure}

In this section, we present the hybrid QGE algorithm with a loop for trial-state optimization, as illustrated in Fig.~\ref{fig_2_flowchart}(a). Further details are provided below.

\subsection{Quantum process} 

In the quantum process [Fig.~\ref{fig_2_flowchart}(b)], each run is iterated over discrete time $t_n = n\delta t$, where $n\in[0,L-1]$, for Fourier sampling in the postprocess. For a single run, $N$ qubits are prepared or reset in the quantum registers, ($q_0$, $q_1$, $\cdots$, $q_{N-1}$), and rotated by the unitary $U_{\rm I}(\vec{\theta})$ to create a trial state with adjustable parameters $\vec{\theta}$ for optimization. Next, the trial state is time-evolved by the first-order Trotterized unitary (higher orders can also be applied)~\cite{Trotter1959,Suzuki1976}:
\begin{equation}
U_M(t_n) = (e^{-iH_1t_n/M}e^{-iH_2t_n/M})^M,
\label{Trotterized_unitary}
\end{equation}
with Trotter depth $M$, which is a compact implementation of the exact time propagator $e^{-iHt_n}$ for a quantum many-body model $H = H_1 + H_2$ with $[H_1,H_2]\neq 0$. Lastly, the output state is rotated back to compensate using $U_{\rm I}(\vec{\theta})$, and measured on the same circuit to return the diagonal components of the time propagator to the classical register, $c$.  The $z$-basis measurement outcomes are represented by:
\begin{equation}
\mathcal{P}_{n} = {\rm Tr}[\rho_0\rho_{M,\vec{\theta}}(t_n)],
\label{propagator_noiseless}
\end{equation}
where the density matrix is defined as $\rho_0 = \prod_{j=0}^{N-1} |0\rangle_j^z \langle 0|_j^z$ for the input registers, and 
\begin{align}
\rho_{M,\vec{\theta}}(t_n) 
& = [U_{\rm I}(\vec{\theta})]^\dag U_M(t_n) U_{\rm I}(\vec{\theta}) \rho_0 [U_{\rm I}(\vec{\theta})]^\dag
\nonumber\\
& ~\times [U_M(t_n)]^\dag U_{\rm I}(\vec{\theta}),
\label{output_noiseless}
\end{align}
for the output state, respectively. Our measurement protocol is similar to ancilla-free variational circuits~\cite{Tilly2022}, but differs from the original QPE circuits~\cite{Kitaev1996,Lloyd1996,Abrams1999} and Hadamard test circuits~\cite{Somma2002}, both of which use controlled measurements via ancilla qubits.

\subsection{Postprocess}
\label{sec_postprocess}

The output states are postprocessed to construct time-series data, $\{(t_n,\mathcal{P}_n)\}_{n=0}^{L-1}$. The time series is filtered by the function $\mathcal{F}_n = \mathcal{F}(t_n)$, satisfying $\mathcal{F}(t_n$ $\rightarrow\infty)\rightarrow 0$, to improve the performance of Fourier sampling and Trotterization~\cite{Lee2024}, and fed into the classical subroutine for a fast Fourier transform (FFT)~\cite{Duhamel1990} that reduces the computational complexity of the discrete Fourier transform (DFT), $\mathcal{O}(L^2)$ to $\mathcal{O}(L\log L)$. In our context, FFT is promoted from DFT yielding a spectral function:
\begin{equation}
\mathcal{A}(\omega_m) = \frac{\delta t}{2\pi} {\rm Re} \sum_{s=\pm} \sum_{n=0}^{L-1} e^{i\omega_m t_{sn}} \mathcal{F}_n \mathcal{P}_{sn},
\label{spectral_function}
\end{equation}
where we define discrete frequencies $\omega_m = m\delta\omega$, conjugate to $t_n$, in units of $\delta t$ and $\delta\omega$ satisfying $\delta\omega \delta t = 2\pi/L$, $m,n \in [0,L-1]$, and $s$ counts the contributions from causal and anti-causal processes. Hereafter, we drop discrete indices from $t_n$ and $\omega_m$ for convenience. 

We can interpret  Eq.~\eqref{spectral_function} in the continuum limit ($L\rightarrow\infty$). It is convenient to recast Eq.~\eqref{spectral_function} in the convolution form:
\begin{gather}
\mathcal{A}(\omega) = \int_{-\infty}^\infty d\tilde{\omega}\tilde{\mathcal{F}}(\tilde{\omega})\mathcal{A}_0(\omega-\tilde{\omega}),
\label{spectral_function_convolution}
\\
\mathcal{A}_0(\omega) = \frac{1}{2\pi} \int_{-\infty}^\infty dt \cos(\omega t) \mathcal{P}(t),
\label{spectral_function_continuum}
\end{gather}
where the filter is defined in Fourier space: $\tilde{\mathcal{F}}(\omega) = \frac{1}{\pi}\int_{0}^\infty dt$ $\cos(\omega t) \mathcal{F}(t)$. In this work, we consider two specific examples with a Lorentzian line shape $\tilde{\mathcal{F}}_{\rm L}(\omega) = \frac{1}{\pi}\frac{\eta}{\omega^2 + \eta^2}$ for $\mathcal{F}_{\rm L}(t) = e^{-\eta t}$ and a Gaussian line shape $\tilde{\mathcal{F}}_{\rm G}(\omega) = \frac{1}{\sqrt{2\pi}\sigma}$ $e^{-\frac{\omega^2}{2\sigma^2}}$ for $\mathcal{F}_{\rm G}(t) = e^{-\sigma^2 t^2/2}$, where $2\eta$ defines the full width at half maximum of the line shape and has a connection to $\sigma$: $\eta = \sigma\sqrt{2\ln 2}$. To proceed, we, for clarity, ignore Trotter truncation error until later to approximate $U_M(t)\approx e^{-iHt}$, and expand the input wavefunction in terms of eigenstates: $|\psi_{\rm I}\rangle = U_{\rm I}(\vec{\theta}) \prod_{j=0}^{N-1} |0\rangle_{j}^z = \sum_u c_u|u\rangle$ with $|u\rangle$ satisfying $H|u\rangle = \mathcal{E}_u|u\rangle$. Substituting Eqs.~\eqref{propagator_noiseless}-\eqref{output_noiseless} into Eqs.~\eqref{spectral_function_convolution}-\eqref{spectral_function_continuum}, we find that the spectral function can be expressed in Lehmann representation:
\begin{equation}
\mathcal{A}(\omega) = \sum_{u,u'} w_{uu'} \tilde{\mathcal{F}}(\omega - \Delta_{uu'}),
\label{noiseless_spectral_function_Lehmann}
\end{equation}
yielding multiple {\it signal} peaks returning exact gaps $\Delta_{uu'} = \mathcal{E}_u - \mathcal{E}_{u'}$ at their centers in association with the spectral weight $w_{uu'} = |c_u|^2|c_{u'}|^2$. Crucially, exact gaps are returned for \emph{any} initial state with non-zero overlap with the exact state of interest, consistent with the Gell-Mann and Low theorem \cite{Gellmann1951}.

Equation~\eqref{spectral_function} and its resulting expression, Eq.~\eqref{noiseless_spectral_function_Lehmann}, establish a mapping between the energy gaps in $H$ and the frequencies of the time propagator $\mathcal{P}_n$. To target a specific gap, an initial estimate can be obtained using classical methods, with the actual gap determined by locating the signal peak closest to this estimate. Although time-series filtering offers advantages, it can introduce errors in gap estimation by distorting the baseline of the signal peak. To address this, baseline correction can be applied using the fitting methods discussed later.

Finally, the trial-state parameters are updated by a classical optimizer to enhance the target signal peak. This approach is distinct from ground-state energy minimization typically employed in variational algorithms. The optimization process proceeds until it meets convergence criteria within a tolerance.

\section{Noise resilience}
\label{sec_noise_resilience}

\begin{figure}[b]
\includegraphics[width=0.4\textwidth]{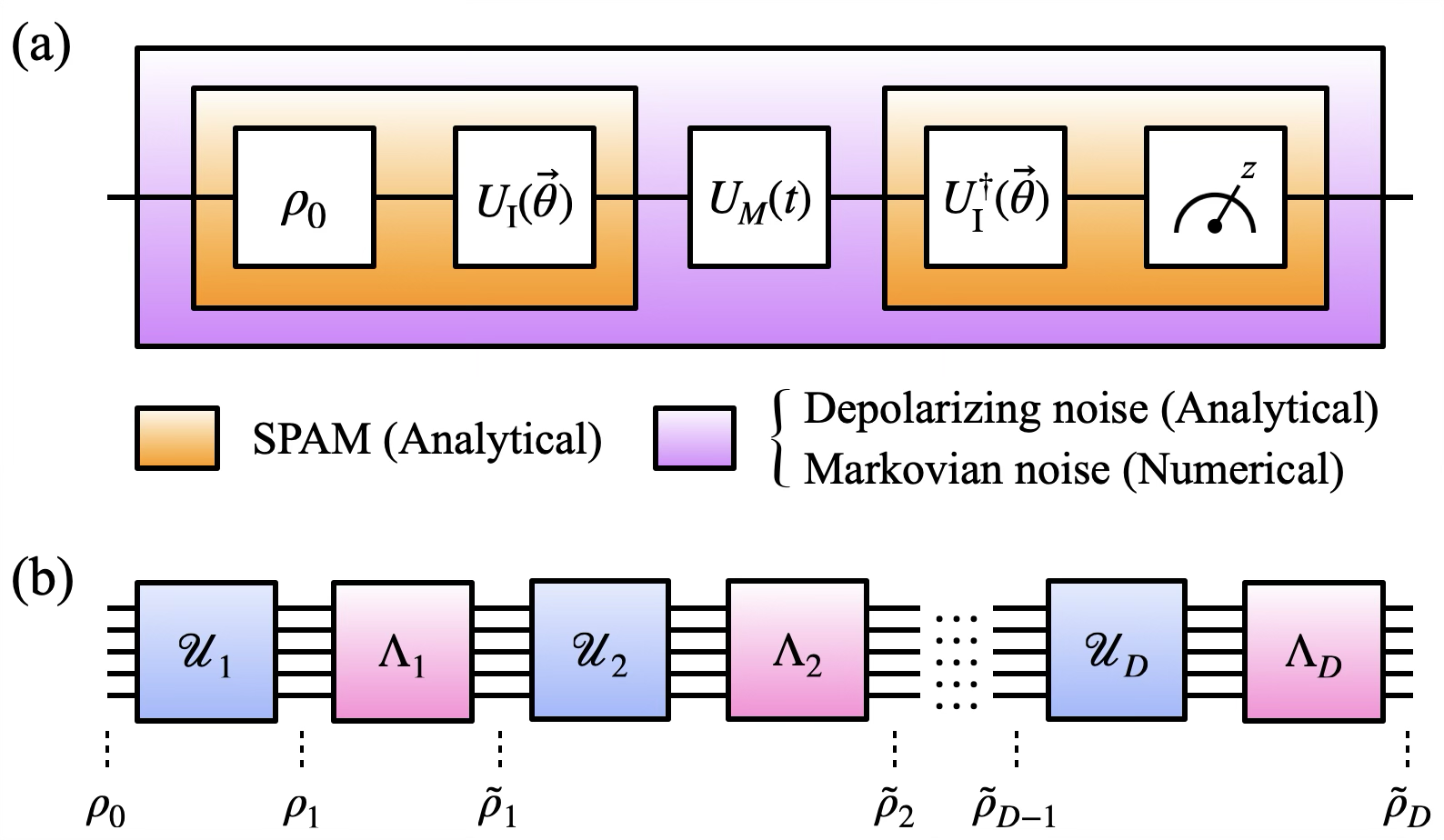}
\caption{(a) Block diagram illustrating the noise types to which each component of a quantum circuit [Fig.~\ref{fig_2_flowchart}(b)] is resilient. The orange shaded area represents resilience to SPAM errors (as analytically demonstrated). The purple shaded area indicates resilience to depolarizing noise (as analytically demonstrated) or Markovian noise (as supported by numerical evidence). The category of Markovian noise encompasses all possible types of multi-qubit depolarizing, bit/phase flips, as well as decoherence and dephasing channels. (b) Block diagram of a general noisy Trotterized circuit with the binary sequence of unitary/noise maps, $(\mathcal{U}_\nu, \Lambda_\nu)$ for $\nu\in[1,D]$.}
\label{fig_3_block_diagram_noise}
\end{figure}

In this section, we analyze the noise resilience of the hybrid QGE algorithm under state preparation and measurement (SPAM) errors and mid-circuit noise. We begin by proving the resilience to SPAM errors, then extend the analysis to a depolarizing-only noise model, and finally broaden the scope to include Markovian noise, accounting for non-depolarizing noise. The key findings are summarized in Fig.~\ref{fig_3_block_diagram_noise}(a).

\subsection{SPAM errors} 

We first prove that our algorithm is resilient to SPAM errors. Starting with the output state, Eq.~\eqref{output_noiseless}, in the absence of errors, we can model state preparation errors by replacing the unitary $U_{\rm I}(\vec{\theta})$ with an alternative form $U'_{\rm I}(\vec{\theta}')$. Similarly, measurement errors can be accounted by replacing the inverse unitary $[U_{\rm I}(\vec{\theta})]^\dag$, applied at the end of the circuit, with $[U''_{\rm I}(\vec{\theta}'')]^\dag$. As a result, the output state affected by SPAM errors reads as:
\begin{align}
\tilde{\rho}_{M,\vec{\theta}',\vec{\theta}''}(t) 
& = [U''_{\rm I}(\vec{\theta}'')]^\dag U_M(t) U'_{\rm I}(\vec{\theta}') \rho_0 [U'_{\rm I}(\vec{\theta}')]^\dag
\nonumber\\
& ~\times [U_M(t)]^\dag U''_{\rm I}(\vec{\theta}'').
\label{output_SPAM_error}
\end{align}
Following the approach used to derive Eq.~\eqref{noiseless_spectral_function_Lehmann}, we perform an eigenstate expansion: $U'_{\rm I}(\vec{\theta}') \prod_{j=0}^{N-1} |0\rangle_{j}^z = \sum_u c'_u|u\rangle$ and $U''_{\rm I}(\vec{\theta}'') \prod_{j=0}^{N-1} |0\rangle_{j}^z = \sum_u c''_u|u\rangle$, where $c'_u\neq c''_u$ in general. Plugging Eq.~\eqref{output_SPAM_error} in Eq.~\eqref{spectral_function}, we find the noisy spectral function:
\begin{equation}
\tilde{\mathcal{A}}_{\rm SPAM}(\omega) = \sum_{u,u'} \tilde{w}_{uu'} \tilde{\mathcal{F}}(\omega - \Delta_{uu'}),
\label{SPAM_spectral_function_Lehmann}
\end{equation}
where the spectral weight is modified as $\tilde{w}_{uu'} = {\rm Re}[(c''_u)^* c'_u$ $(c'_{u'})^* c''_{u'}]$.

We conclude that the sole impact of SPAM errors in Eq.~\eqref{SPAM_spectral_function_Lehmann} is a modification of the spectral weights, while the peak locations remain unchanged. SPAM errors alone do not produce erroneous {\it satellite} peaks or cause shifts in signal peaks (as long as they are well-separated, typically by more than $2\eta$). This confirms that the hybrid QGE algorithm is inherently resilient to SPAM errors. Moreover, any peak suppression due to SPAM errors can be mitigated through trial-state optimization, as discussed in Sec.~\ref{sec_numerical_results}.

\subsection{Depolarizing noise}
\label{sec_depolarizing}

Next, we prove that the hybrid QGE algorithm returns \emph{exact} energy gaps in the presence of {\it multi-qubit} depolarizing noise. The proof relies on the fact that $e^{-iHt}$ applied to an \emph{arbitrary} initial state oscillates in time at frequencies matching the exact gaps of $H$ \cite{Lee2024}. To begin with, we consider a full circuit-level noise model that allows arbitrary types of noise to arise at any stage in the quantum processes [see Fig.~\ref{fig_3_block_diagram_noise}(b)]. Ignoring temporal correlations (memory) between processes at different stages, thus maintaining the Markovianity of the processes~\cite{Breuer2016}, the noisy output state of a quantum circuit with total depth $D$ can be described by the density matrix:
\begin{equation}
\tilde{\rho}_D = (\Lambda_D\circ\mathcal{U}_D\circ\cdots\circ\Lambda_2\circ\mathcal{U}_2\circ\Lambda_1\circ\mathcal{U}_1)(\rho_0),
\label{noisy_density_matrix}
\end{equation}
where, for $\nu\in[1,D]$, the unitary map $\mathcal{U}_\nu$ evolves $\rho$ to $U_\nu \rho U_\nu^\dag$ with the unitary $U_\nu$ for either initialization or time evolution, and the (completely positive trace preserving) noise map $\Lambda_\nu$ describes the non-unitary evolution of a quantum state by a certain type of noise channel. 

\begin{figure}[t]
\begin{center}
\includegraphics[width=0.48\textwidth]{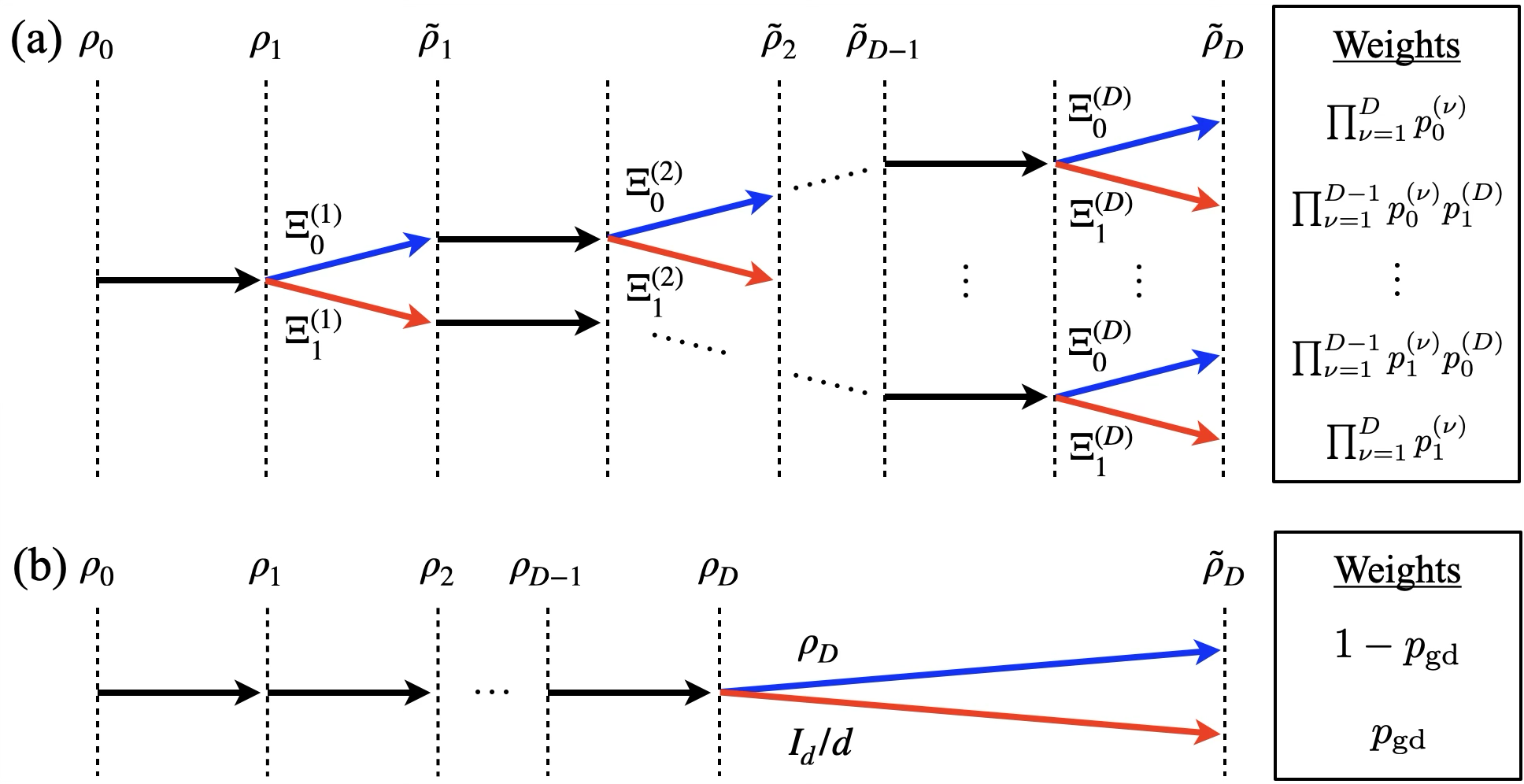}
\end{center}
\vspace{-15pt}
\caption{(a) Branching diagram showing all contributions to the noisy output density matrix $\tilde{\rho}_D$ arising from $\Lambda_\nu$ for the mid-circuit multi-qubit depolarizing channel [Eq.~\eqref{depolarizing_map}]. In each block, corresponding to Fig.~\ref{fig_3_block_diagram_noise}(b), the black arrow indicates unitary evolution, and the blue (or red) arrow describes an idle process (or non-unitary evolution) with the probability $p_0^{(\nu)}$ (or $p_1^{(\nu)}$). The associated weights for available branches are listed in the right panel. (b) Branching diagram for the equivalent block diagram to Fig.~\ref{fig_3_block_diagram_noise}(b) with the noise map $\Lambda_{\rm gd}$ for global depolarizing channel [Eq.~\eqref{noisy_density_matrix_depolarizing}].}
\label{fig_4_branchingdiagram_depolarizing}
\end{figure}

For the $N$-qubit depolarizing channel, a compact analytic result can be derived. Since all branches of quantum processes cooperate to form a maximally mixed state~\cite{Comment_depolarizing}, the noise map is significantly simplified into:
\begin{equation}
\Lambda_\nu(\rho) = p_0^{(\nu)} \Xi_0^{(\nu)}(\rho) + p_1^{(\nu)} \Xi_1^{(\nu)}(\rho),
\label{depolarizing_map}
\end{equation}
where we define $\Xi_0^{(\nu)}(\rho) = \rho$, $\Xi_1^{(\nu)}(\rho) = I_d/d$, $I_d = {\rm diag}(1,$ $1,\cdots,1)$ of size $d = 2^N$, and $p_1^{(\nu)}$ is the depolarizing probability at stage $\nu$, which satisfies $p_0^{(\nu)} + p_1^{(\nu)} = 1$. Leveraging the structure of Eq.~\eqref{depolarizing_map}, we can count all contributions to Eq.~\eqref{noisy_density_matrix} with the associated weights [Fig.~\ref{fig_4_branchingdiagram_depolarizing}(a)]. We then recast Eq.~\eqref{noisy_density_matrix} into the equivalent form [Fig.~\ref{fig_4_branchingdiagram_depolarizing}(b)]:
\begin{align}
\tilde{\rho}_D 
& = (\Lambda_{\rm gd}\circ\mathcal{U}_D\circ\cdots\circ\mathcal{U}_2\circ\mathcal{U}_1)(\rho_0)
\nonumber\\
& = (1 - p_{\rm gd}) (\mathcal{U}_D\circ\cdots\circ\mathcal{U}_2\circ\mathcal{U}_1)(\rho_0) + p_{\rm gd} I_d/d,
\label{noisy_density_matrix_depolarizing}
\end{align}
where $\Lambda_{\rm gd}$ defines the noise map for {\it global} depolarizing channel with the probability $p_{\rm gd} = 1 - \prod_{\nu=1}^{D}(1-p_1^{(\nu)})$. For the specific setup in Fig.~\ref{fig_2_flowchart}(b), we can rearrange: $U_D\cdots U_2U_1 = [U_{\rm I}(\vec{\theta})]^\dag U_M(t)U_{\rm I}(\vec{\theta})$. Note that Eq.~\eqref{noisy_density_matrix_depolarizing} has been considered in various contexts including robust data encoding for quantum classifiers~\cite{LaRose2020}, error mitigation strategies~\cite{Vovrosh2021, Urbanek2021}, and Grover's algorithm~\cite{Stoudenmire2024}. Plugging Eq.~\eqref{noisy_density_matrix_depolarizing} in Eq.~\eqref{spectral_function} along with Eq.~\eqref{propagator_noiseless} and expanding $U_M(t)$ to first order in $M^{-1}$, we derive the noisy spectral function: 
\begin{align}
\tilde{\mathcal{A}}_{\rm gd}(\omega) 
& = (1 - p_{\rm gd}) \big[ \mathcal{A}(\omega) + M^{-1} \delta\mathcal{A}(\omega) + \mathcal{O}(M^{-2}) \big]
\nonumber\\
& ~~~ + p_{\rm gd} \tilde{\mathcal{F}}(\omega)/d.
\label{noisy_spectral_function_depolarizing}
\end{align}
In the noiseless part, the leading term $\mathcal{A}$ [Eq.~\eqref{noiseless_spectral_function_Lehmann}] represents the signal peaks, while the subleading term $M^{-1}\delta\mathcal{A}$ (see Appendix~\ref{appendix_derivation_depolarizing}), arising from Trotter truncation error, accounts for the satellite peaks. The last term reflects the impact of the global depolarizing channel. 

Equation~\eqref{noisy_spectral_function_depolarizing} presents one of the main results in this paper and demonstrates the algorithm's inherent resilience to depolarizing noise. We note several important observations: 

(i) {\it Background formation}: The last term in Eq.~\eqref{noisy_spectral_function_depolarizing}, along with the term for $u = u'$ in Eq.~\eqref{noiseless_spectral_function_Lehmann}, introduces a redundant peak at $\omega = 0$, which defines the lower bound of the target window for QGE. This peak serves as a background, characterized by a broadening parameter $\eta$ and a scaled height of $1 - (1 - 1/d) p_{\rm gd}$.

(ii) {\it Baseline correction}: Both the signal and satellite peaks are exponentially suppressed as $D$ increases such that $1-p_{\rm gd} \approx e^{-Dp_{\rm dp}}$ for $p_{\rm dp} = p_1^{(\nu)} \ll 1$. Assuming that $Dp_{\rm dp}$ is bounded, this issue can be addressed by correcting the baseline distorted by the background using standard fitting methods. Here, Lorentzian or Gaussian fitting works if the redundant peak is well separated from surrounding peaks, typically by more than $2\eta$.

(iii) {\it Trotter depth's robustness to depolarizing noise}: The baseline-corrected signal peaks yield exact gaps at their centers if the satellite peaks in $\delta\mathcal{A}$ are sufficiently suppressed for $M \gg M_c$. The Trotter depth cutoff, $M_c$, for a fixed tolerance $\varepsilon_{{\rm T},c}$, is given by~\cite{Lee2024}:
\begin{equation} M_c = \frac{||[H_1,H_2]||}{\varepsilon_{{\rm T},c}} t^2 \mathcal{F}(t),
\label{Trotter_depth_cutoff}
\end{equation}
where $M_c$ improves exponentially with increasing $\eta$, albeit at the expense of reduced spectral resolution. A key finding is that $M_c$ remains unaffected by $p_{\rm gd}$ due to the commutativity of the Hamiltonian with the depolarizing noise map $\Lambda_{\rm gd}$.

(iv) {\it Targeting a signal peak}: To focus on a specific gap within the target window, it is crucial to distinguish the target signal peak from surrounding peaks. Different configurations of $U_{\rm I}(\vec{\theta})$ can amplify specific peak heights, as represented by $c_u = \bra{u} U_{\rm I}(\vec{\theta}) \prod_{j=0}^{N-1} |0\rangle_{j}^z$ in Eq.~\eqref{noiseless_spectral_function_Lehmann}. To demonstrate the advantage of quantum simulation using our algorithm, trial-state optimization must enhance the signal peak height, making it observable for circuit widths (or qubit counts) that are otherwise infeasible for classical machines due to memory constraints.

(v) {\it Practical relevance}: Although our noise resilience argument focuses on depolarizing noise, it has significant practical relevance for several reasons. First, the depolarizing-only noise model has been experimentally validated for trapped-ion quantum computers, indicating that depolarizing noise is a dominant error source in these systems~\cite{Postler2022, Heussen2023}. Second, other types of noise, including coherent noise channels, can be converted into incoherent depolarizing noise by applying random single-qubit gates, a technique known as twirling~\cite{Magesan2011, Wallman2016, Cai2019, Cai2020, Hangleiter2023}.

\subsection{Markovian noise}
\label{sec_Markovian}

Until now, we have focused on SPAM errors and  depolarizing noise to demonstrate the algorithm's inherent noise resilience and explored how to address other noise types within the same framework. In fact, the noise resilience argument can be extended to include Markovian noise, accounting for non-depolarizing noise, while it is generally challenging. In our approach, we construct an equivalent noisy quantum circuit [in the same spirit as Eq.~\eqref{noisy_density_matrix_depolarizing}] and develop an error mitigation method solely based on classical signal processing techniques.

We start by considering a setup where all mid-circuit layers are subject to both depolarizing and non-depolarizing channels. In particular, for transmon-based quantum computers, non-depolarizing channels include bit/phase flips during state preparation and measurement, as well as decoherence and dephasing caused by interactions with the environment~\cite{Georgopoulos2021}. It is convenient to introduce the general noise map describing all types of Markovian noise~\cite{Siudzinska2020}:
\begin{equation}
\Lambda_\nu(\rho) = \sum_{k_\nu=0}^{d+1} p_{k_\nu}^{(\nu)} \Xi_{k_\nu}^{(\nu)}(\rho),
\label{Generalized_Pauli_map}
\end{equation}
where the maps for $d+2$ branches ($d = 2^N$) are defined as: $\Xi_0^{(\nu)}(\rho) = \rho$, and
\begin{equation}
\Xi_{k_\nu\geq 1}^{(\nu)}(\rho) = \frac{1}{d-1} \sum_{q_\nu=1}^{d-1} K_{k_\nu q_\nu}^{(\nu)}\rho[K_{k_\nu q_\nu}^{(\nu)}]^\dag,
\label{Generalized_Pauli_map_branches}
\end{equation}
with the (normalized) Kraus operators $K_q = \sum_{l=0}^{d-1} e^{2\pi i ql/d}$ $|\psi_q\rangle\langle\psi_q|$ built from the basis state $\{|\psi_q\rangle\}$, and $p_{k_\nu}^{(\nu)}$ is the noise probability satisfying $\sum_{k_\nu=0}^{d+1} p_{k_\nu}^{(\nu)} = 1$. It appears that addressing Eq.~\eqref{noisy_density_matrix} becomes challenging when $[\mathcal{U}_\nu, \Lambda_{\nu'}] \neq 0$.

To properly address the issue of non-commuting maps, we tackle the problem by constructing an equivalent noisy quantum circuit to Eq.~\eqref{noisy_density_matrix}, i.e., by successively swapping $\mathcal{U}_\nu$ and $\Lambda_{\nu'}$ via Pauli propagation and pushing all $\Lambda_\nu$'s to the circuit end (see Appendix~\ref{appendix_Pauli_propagation}). Plugging the result in Eq.~\eqref{spectral_function}, in conjuction with Eq.~\eqref{propagator_noiseless}, and expanding $U_M(t)$ to the first order in $M^{-1}$, we can derive the noisy spectral function:
\begin{align}
& \tilde{\mathcal{A}}_{\rm gp}(\omega) = p_{\vec{0}} \big[ \mathcal{A}(\omega) + M^{-1} \delta\mathcal{A}(\omega) + \mathcal{O}(M^{-2}) \big]
\nonumber\\
& ~~ + \sum_{\vec{k}\neq\vec{0}} p_{\vec{k}} \big[ \tilde{\mathcal{A}}_{\rm gp}^{(\vec{k})}(\omega) + M^{-1} \delta\tilde{\mathcal{A}}_{\rm gp}^{(\vec{k})}(\omega) + \mathcal{O}(M^{-2}) \big],
\label{noisy_spectral_function_nondepolarizing}
\end{align}
where the noiseless terms, $\mathcal{A}$ and $\delta\mathcal{A}$, are defined in the same ways as in Eq.~\eqref{noisy_spectral_function_depolarizing}, but with the adjusted probability $p_{\vec{0}} = \prod_{\nu=1}^D p_0^{(\nu)}$, and $\tilde{\mathcal{A}}_{\rm gp}^{(\vec{k})}$ is the leading correction term by noise channels with the probability $p_{\vec{k}} = \prod_{\nu=1}^D p_{k_\nu}^{(\nu)}$, satisfying $\sum_{\vec{k}} p_{\vec{k}} = 1$, for the choice of $\vec{k} = (k_1,k_2,\cdots,k_D)$:
\begin{align}
\tilde{\mathcal{A}}_{\rm gp}^{(\vec{k})}(\omega) 
& = \sum_{u,u'} \sum_{r_1,r'_1} \sum_{r_2,r'_2} \tilde{w}_{uu'r_1r'_1r_2r'_2}^{(\vec{k})}
\nonumber\\
& ~\times \tilde{\mathcal{F}}(\omega - \Delta_{uu'} - \delta\Delta_{r_1r'_1r_2r'_2}),
\label{spectral_function_noisy_correction}
\end{align}
describing satellite peaks with centers shifted from $\Delta_{uu'}$ by 
\begin{align}
\delta\Delta_{r_1r'_1r_2r'_2} = 2M^{-1}[(r_1 - r'_1) h_1 + (r_2 - r'_2) h_2],
\label{gap_nondepol}
\end{align}
and the associated spectral weights:
\begin{equation}
\tilde{w}_{uu'r_1r'_1r_2r'_2}^{(\vec{k})} = \frac{1}{(d-1)^D} \sum_{\vec{q}} {\rm Re} \big\{ [\tilde{c}_{ur_1r_2}^{(\vec{k}\vec{q})}]^* c_u c^*_{u'} \tilde{c}_{u'r'_1r'_2}^{(\vec{k}\vec{q})} \big\}.
\label{spectral_weight_nondepol}
\end{equation}
In the above expression, $\tilde{c}_{ur_1r_2}^{(\vec{k}\vec{q})}$ is defined in Appendix~\ref{appendix_derivation_nondepolarizing}, and $r_{1,2}$, $r'_{1,2}$ denote the Fourier modes in the double series expansion of $\tilde{K}(t)$. The subleading term $M^{-1}\delta\tilde{\mathcal{A}}_{\rm gp}^{(\vec{k})}$ (see Appendix~\ref{appendix_derivation_nondepolarizing}) describes Trotter truncation error for each choice of noise channel $\vec{k}\neq\vec{0}$. 

Equations~\eqref{noisy_spectral_function_nondepolarizing}-\eqref{spectral_weight_nondepol} provide a foundation for further exploration of noise resilience in a more general setting. In contrast to the depolarizing case, the corrections by non-depolarizing noise are not simply formulated, making it challenging to derive insights from the studies on individual noise channels. In our approach, we focus on Eq.~\eqref{spectral_function_noisy_correction}, along with Eqs.~\eqref{gap_nondepol}-\eqref{spectral_weight_nondepol}, that captures the leading behavior of noise-induced satellite peaks. Several key observations emerge: 

(i) {\it Background formation}: In the non-depolarizing case, a complex interplay exists between the satellite peak shift $\delta\Delta$ and Trotter depth $M$. The $M^{-1}$-dependence in Eq.~\eqref{gap_nondepol} indicates that noise-induced satellite peaks are distributed across the frequency domain with relatively small peak-to-peak separations for $M\gg M_c$. For ${\rm max}\{|h_1|,|h_2|\}$ $\ll M\eta$, however, individual satellite peaks lose their resolution, merging into a slowly varying background. For each choice of $\vec{k}$, this noise-induced background serves as a potential error source in gap estimation, particularly when it varies rapidly around the target signal peak, causing drift from the exact gap. 

(ii) {\it Cumulative effect of noise channels}: Background formation can be fully characterized by accounting for all noise channels. The height of each satellite peak is suppressed by $p_{\vec{k}}$, which becomes exponentially small for large $M$, assuming the probability of each individual noise channel, $p_{k_{\nu}}$, is low. This presents a challenge since the diminishing height of individual peaks could be difficult to track. However, the cumulative effect can still become significant due to the large number of terms in the summation ($\sim 2^{ND} \gg M$) in Eq.~\eqref{noisy_spectral_function_nondepolarizing}. Consequently, it is reasonable to establish a threshold value for $p_{\vec{k}}$, above which the cumulative contribution of $\sum_{\vec{k}\neq\vec{0}} p_{\vec{k}} \tilde{\mathcal{A}}_{\rm gp}^{(\vec{k})}(\omega)$ leads to a noise-induced background that overwhelms the signal peaks arising from $p_{\vec{0}}\mathcal{A}(\omega)$. The specific threshold value depends on the structure of the quantum circuit, the types of noise channels acting on it, and the degree to which the noise maps commute with the unitary maps comprising the circuit.

(iii) {\it Breakdown of positive definiteness}: Unlike the noiseless case [Eq.~\eqref{noiseless_spectral_function_Lehmann}], the noise correction [Eq.~\eqref{spectral_function_noisy_correction}] is not guaranteed to be positive definite. This occurs because the spectral weight [Eq.~\eqref{spectral_weight_nondepol}] is not bounded below by zero, highlighting the impact of non-depolarizing noise. Additionally, combinations of $r_{1,2}$ and $r'_{1,2}$, that lead to large shifts in Eq.~\eqref{gap_nondepol} are generally disfavored, as peaks centered at higher frequencies have diminishing spectral weights in finite-size systems. 

(iv) {\it Interplay with Trotter truncation error}: The subleading term, $M^{-1}\delta\tilde{\mathcal{A}}_{\rm gp}^{(\vec{k})}$, arises from Trotter truncation error under non-depolarizing noise and is suppressed for $M \gg M_c$ relative to the leading term with the same index $\vec{k}$. This suppression mirrors the behavior observed in the noiseless case.  

(v) {\it Baseline correction}: To restore signal peaks to original locations and improve the algorithm’s robustness to noise, effective background removal is crucial. Building on advancements in classical data processing for spectroscopy~\cite{Gautam2015}, we propose an error mitigation strategy that leverages exclusively classical signal processing techniques. Notably, this approach avoids any additional overhead associated with qubits or gate operations, offering a distinct advantage over conventional methods~\cite{Cai2023}. Specifically, we choose the asymmetric least squares (ALS) fitting method for baseline correction of noisy spectral data~\cite{Eilers2005}, along with other variations~\cite{Zhang2010, He2014, Baek2015, Korepanov2020}. See Appendix~\ref{appendix_ALS} for further details.

\section{Numerical results}
\label{sec_numerical_results}

In this section, we present numerical demonstrations of the algorithm’s procedure, along with its robustness to noise and scalability, using a minimal spin model.

\begin{figure*}[t]
\begin{center}
\includegraphics[width=0.96\textwidth]{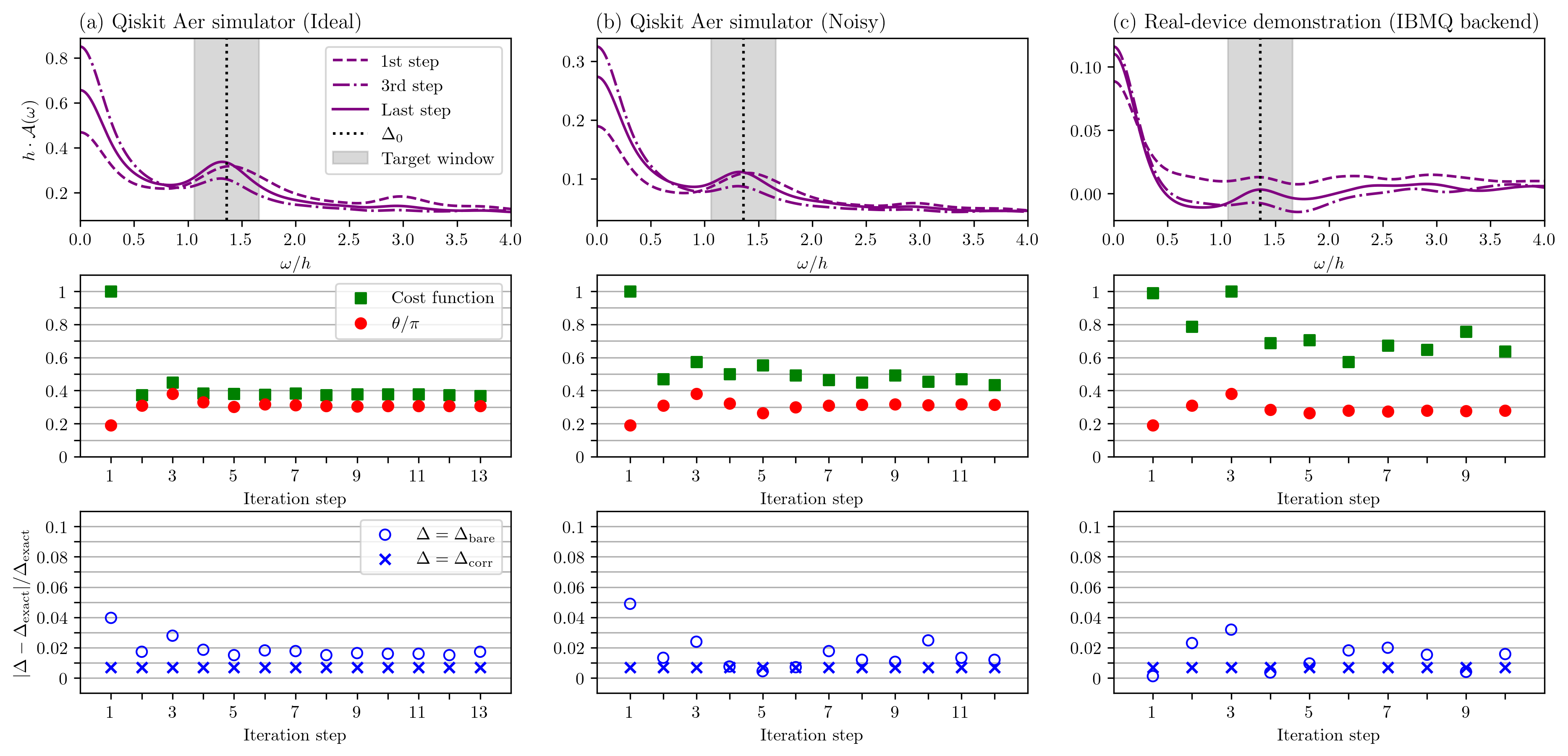}
\end{center}
\vspace{-15pt}
\caption{Simulation results demonstrating the trial-state optimization process in QGE using different approaches: (a) Qiskit Aer simulator without a noise model, (b) Qiskit Aer simulator with a noise model calibrated from the IBM Quantum backend, {\it ibm\_sherbrooke}, and (c) real-device demonstration on the same backend. Top panels: Spectral function $\mathcal{A}$ versus $\omega$ for the TFIM of $N=5$, $J/h=0.4$ and the filter $\mathcal{F}=\mathcal{F}_{\rm L}$ with $\eta/h=0.3$, shown at the first (dashed curve), intermediate (dot-dashed curve) and last (solid curve) iteration steps. The bare gap $\Delta_{\rm bare}$ is estimated by locating the signal peak closest to the initial gap $\Delta_0$ (black dotted vertical line). By setting the target window $\Delta_0-\eta\leq\omega\leq\Delta_0+\eta$ (shaded in gray) and correcting the baseline of the spectral function (see Appendix~\ref{appendix_QGE_procedure} for technical details), the corrected gap $\Delta_{\rm corr}$ is estimated in the same way as $\Delta_{\rm bare}$. Middle and bottom panels: Progress of the input orientation (red filled circles), the cost function (defined as the normalized inverse height of the target signal peak at $\omega = \Delta_{\rm corr}$) (green filled squares), and the gap estimate error $|\Delta - \Delta_{\rm exact}| / \Delta_{\rm exact}$ for $\Delta = \Delta_{\rm bare}$ (blue open circles) and $\Delta = \Delta_{\rm corr}$ (cross markers), respectively, as a function of the iteration step. Here, a variant of Brent’s method is employed for single-variable bounded optimization with tolerance $10^{-6}$. The iterations are terminated early for demonstration purposes. In all simulations, 1024 measurement shots are commonly used with the parameters: $M=15$, $\delta\omega = \eta/4$, $L = 2\lceil 5h/\delta\omega\rceil$, where $\lceil x\rceil$ is the ceiling function of $x$.}
\label{fig_5_optimization_N5_FL}
\end{figure*}

\subsection{Simulation setup}

For clarity in demonstrating our algorithm, we choose the transverse-field Ising model (TFIM) in one dimension with open boundaries: 
\begin{equation}
H_1 = - J \sum_{j=0}^{N-2}\sigma^z_j \sigma^z_{j+1},
~H_2 = - h \sum_{j=0}^{N-1} \sigma^x_j,
\label{model_TFIM}
\end{equation}
where $(\sigma^x,\sigma^y,\sigma^z)$ are the Pauli matrices, $j$ indexes spins at sites $j\in[0,N-1]$, $J$ is the Ising coupling, and $h$ is the magnetic field. For quantum circuit implementation [Fig.~\ref{fig_2_flowchart}(b)], the Trotterized time-evolution unitary is specified by: 
\begin{equation}
U_M(t) = \bigg[ \prod_{j=0}^{N-2} R_{j,j+1}^{zz}\bigg(\frac{-2Jt}{M}\bigg) \prod_{j'=0}^{N-1} R_{j'}^x \bigg(\frac{-2ht}{M}\bigg) \bigg]^M,
\label{Trotterized_unitary}
\end{equation}
where we define one and two-qubit rotations as $R_j^{\alpha_j}(\phi) = \exp(-i\frac{\phi}{2}\sigma_j^{\alpha_j})$, $R_{j,j+1}^{\alpha_j\alpha_{j+1}}(\phi) = \exp(-i\frac{\phi}{2}\sigma_j^{\alpha_j}\sigma_{j+1}^{\alpha_{j+1}})$, respectively. Meanwhile, the trial-state unitary is chosen in the form:
\begin{equation}
U_{\rm I}(\vec{\theta}) = \prod_{j=0}^{N-1} R_{j}^y(\theta),
\label{trial_state_unitary}
\end{equation}
with a single parameter $\theta$. Our choice of $U_{\rm I}(\vec{\theta})$ scales linearly with $N$, unlike typical variational ansatzes~\cite{Tilly2022}, as its role is restricted to signal enhancement via optimization.

In our simulations, the primary goal is to estimate the first gap $\Delta_{\rm exact} \equiv \mathcal{E}_2 - \mathcal{E}_1$ in the quantum paramagnetic regime ($J<h$) of a small TFIM ($N<10$) (while other gaps can be estimated in a similar way), demonstrating the algorithm's procedure and noise resilience. The simulations are conducted using different approaches: Qiskit Aer simulator without/with a noise model, calibrated from the IBM Quantum backend, {\it ibm\_sherbrooke}, and real-device demonstrations on the same backend. See Appendix~\ref{appendix_qiskit_overview} for the Qiskit code overview and Appendix~\ref{appendix_backend_calibration} for the backend calibration data. All simulation results in this section and Appendix~\ref{appendix_QGE_procedure} are generated using the Lorentzian filter $\mathcal{F}_{\rm L}$, which can be compared with the Gaussian filter $\mathcal{F}_{\rm G}$ (see Appendix~\ref{appendix_Gaussian_filter}).

\subsection{Algorithm's robustness to noise}
\label{sec_robustness_to_noise}

The spectral function, which is central to our simulations, reveals multiple signal peaks returning gaps at their centers. To extract the information on a target gap from the spectral function, a systematic procedure is required, as outlined in Sec.~\ref{sec_postprocess}. The top panels in Fig.~\ref{fig_5_optimization_N5_FL} illustrate the spectral function’s line shape for $N=5$ and $J/h=0.4$, along with selected values of $\theta$ corresponding to intermediate steps of trial-state optimization. For $\eta/h = 0.3$, a small Trotter depth of $M = 15$ is sufficient~\cite{Lee2024}. To estimate the first gap, we start with an initial guess, $\Delta_0/h = 2[1-(1-1/N)J/h]$, derived by perturbation theory for a quantum paramagnet~\cite{Sachdev2011}. Ideally, the bare gap $\Delta_{\rm bare}$ is estimated by locating the signal peak closest to $\Delta_0$. In practice, however, as the target peak shifts from the exact gap due to baseline distortion from various error sources discussed in Sec.~\ref{sec_noise_resilience}, a robust strategy is needed to accurately restore the peak location. By setting the target window $\Delta_0-\eta\leq\omega\leq\Delta_0+\eta$, baseline distortion can be corrected using classical signal processing techniques. As before, the corrected gap $\Delta_{\rm corr}$ is estimated by locating the signal peak closest to $\Delta_0$. See Appendix~\ref{appendix_QGE_procedure} for further technical details and extended data.

If $\theta$ is suboptimal for any reason, the effectiveness of the algorithm may be compromised, potentially causing the target signal peak to fall below the device-specific detection threshold. To improve the algorithm’s robustness, $\theta$ can be utilized as an optimization variable for the trial state. Following baseline correction, we define the cost function as the inverse height of the target signal peak, which is minimized through trial-state optimization. The middle and bottom panels illustrate the progress of $\theta$, the cost function, and the gap estimate error $|\Delta-\Delta_{\rm exact}|/\Delta_{\rm exact}$ for $\Delta\in\{\Delta_{\rm bare},\Delta_{\rm corr}\}$ as a function of the iteration step. In the noiseless results, we observe that as the iterations progress, the target signal peak converges more closely to $\omega = \Delta_{\rm exact}$, while its relative height increases compared to the redundant peak at $\omega = 0$. Furthermore, baseline correction reduces the gap estimate error to below 1\%. In both the noisy simulation and real-device demonstration, trial-state optimization proves highly effective. Remarkably, baseline correction allows us to achieve a gap estimate error comparable to the noiseless case, highlighting the algorithm's robustness across various noise channels.

\begin{figure*}[t]
\begin{center}
\includegraphics[width=0.92\textwidth]{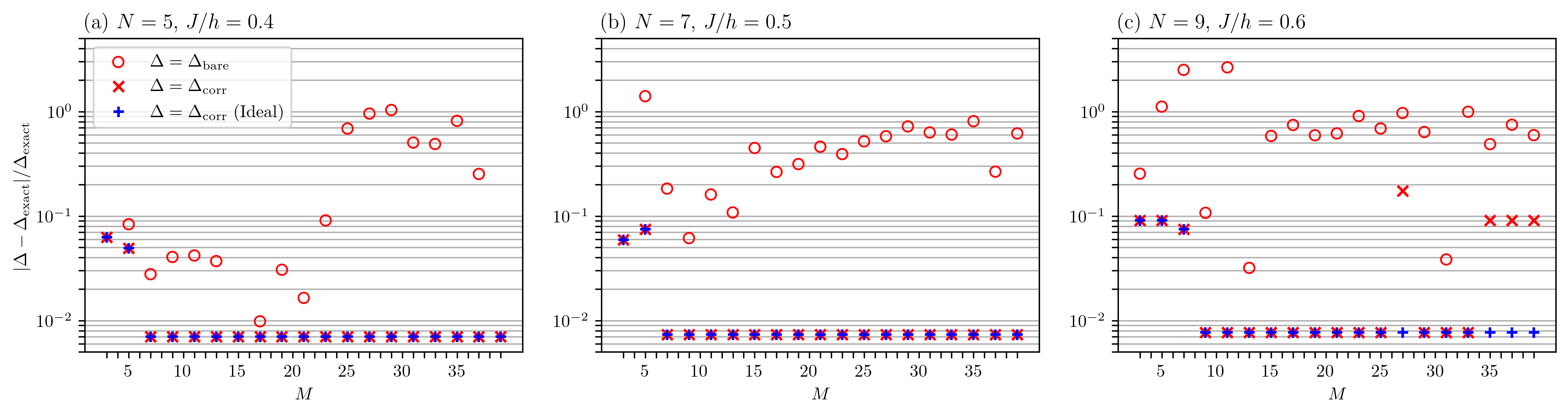}
\end{center}
\vspace{-15pt}
\caption{Demonstrating the scalability and robustness of the hybrid QGE algorithm. Each panel displays the gap estimate error $|\Delta - \Delta_{\rm exact}|$ $/\Delta_{\rm exact}$ as a function of the Trotter circuit depth $M$ for different parameter settings: $(N,J/h) =$ (a) $(5,0.4)$, (b) $(7,0.5)$, (c) $(9,0.6)$. Red symbols represent real-device demonstrations for $\Delta\in\{\Delta_{\rm bare},\Delta_{\rm corr}\}$, while the blue plus marker denotes the reference data obtained from noiseless simulations, as specified in the legend. Here, the vertical axis uses a logarithmic scale, the input orientation is set to the unoptimized value $\theta = 0.3\pi$, and other parameter settings are the same as in Fig.~\ref{fig_5_optimization_N5_FL}.}
\label{fig_6_convergence_FL_compact}
\end{figure*}

\subsection{Algorithm's robustness to scalability}
\label{appendix_circuit_scaling}

Our numerical simulations thus far have focused on a specific parameter setting with a small circuit depth $M=15$ and width $N=5$. To further benchmark the algorithm, it is crucial to investigate its robustness as the system scales. Here, we focus on data for real-device demonstrations (see Appendix~\ref{appendix_QGE_procedure} for extended data). Figure~\ref{fig_6_convergence_FL_compact} illustrates how the gap estimate error $|\Delta - \Delta_{\rm exact}|/\Delta_{\rm exact}$ evolves with increasing $M$. Each panel corresponds to distinct parameter setting for $N$ and $J/h$. 

The red open circle represents the gap estimate error for $\Delta = \Delta_{\rm bare}$ (before baseline correction), highlighting that the algorithm struggles to converge as $M$ and $N$ increase under realistic noise. In the Markovian noise model, which includes non-depolarizing channels [Eq.~\eqref{Generalized_Pauli_map}], this behavior is primarily attributed to the term $\sum_{\vec{k}\neq\vec{0}}p_{\vec{k}}\tilde{\mathcal{A}}_{\rm gp}^{(\vec{k})}$ in Eq.~\eqref{noisy_spectral_function_nondepolarizing}, along with its subleading corrections. As discussed in Sec.~\ref{sec_Markovian}, this term introduces a background around the target signal peak, causing the peak to drift away from its exact location. Increasing $M$ exacerbates this issue by significantly amplifying the background [Eq.~\eqref{gap_nondepol}], which is generally not positive definite [Eq.~\eqref{spectral_weight_nondepol}]. Furthermore, other factors may come into play beyond the assumption of Markovianity~\cite{Breuer2016}, potentially invalidating Eq.~\eqref{noisy_density_matrix}. 

The red cross marker highlights the algorithm's success after baseline correction, demonstrating a significant reduction in the gap estimate error to below 1\%, consistent with the reference data from noiseless simulations (blue plus marker). A few exceptions [red cross markers with errors around 10\% in Fig.~\ref{fig_6_convergence_FL_compact} (c)] can be further improved through trial-state optimization. It is noticeable that residual errors below 1\% persist due to the assumptions of ALS fitting (see Appendix~\ref{appendix_ALS}). While advanced ALS fitting methods~\cite{Zhang2010, He2014, Baek2015, Korepanov2020} or alternative signal processing techniques~\cite{Gautam2015} may potentially eliminate these errors, their resolution is deferred to future work. These findings highlight the potential of our algorithm to contribute to quantum advantage as the system scales, even without relying on active error corrections~\cite{Tehral2015}.

\section{Discussion}
\label{sec_discussion}

We developed a hybrid QGE algorithm that uniquely combines iterative trial-state optimization and classical signal processing techniques for time-series filtering and baseline correction of spectral data, enabling unbiased gap estimates for many-body quantum systems. We analytically demonstrated that the algorithm is inherently resilient to SPAM errors and depolarizing noise. Without incurring additional quantum resources for error mitigation or correction, we provided numerical evidence of its robustness to arbitrary Markovian noise channels, through noisy simulations using the Qiskit Aer simulator and demonstrations on IBM Quantum processors. Our numerical results for a minimal spin model showed that, even in the presence of noise, trial-state optimization and classical signal processing significantly enhance the signal peak at the target gap, thereby effectively reducing the gap estimate error.

These results pave the way for numerous future research directions and applications. For various quantum devices and many-body models of interest, our algorithm can be tested to determine the maximally accessible system size $N_c$, and suggest strategies for scaling to larger $N_c$ values without active error correction. Further analysis could investigate the effects of realistic non-Markovian noise with memory~\cite{Breuer2016}. Furthermore, the noise-resilient framework of our algorithm could be extended to estimate additional correlation functions associated with key physical observables beyond energy gaps and to support other types of hybrid quantum algorithms incorporating time-series analogs. While this is straightforward for ancilla-free circuits, further investigation is required for cases involving controlled operations with ancilla qubits. 

We thank A. F. Kemper, P. Roushan, P. Alsing, S. Patel, and D. Koch for their insightful discussions. This research was supported by the Army Research Office (ARO) under Grant No.~W911NF2210247, and the Air Force Office of Scientific Research (AFOSR) under Grant No.~FA2386-21-1-4081, FA9550-19-1-0272, and FA9550-23-1-0034. The authors acknowledge the Air Force Research Laboratory (AFRL) for support in providing access to IBM Quantum resources. The views expressed are those of the authors, and do not reflect the official policy or position of IBM or the IBM Quantum team.

\appendix

\section{Derivation of the subleading term in Eq.~\eqref{noisy_spectral_function_depolarizing}}
\label{appendix_derivation_depolarizing}

\begin{figure}[b]
\begin{center}
\includegraphics[width=0.48\textwidth]{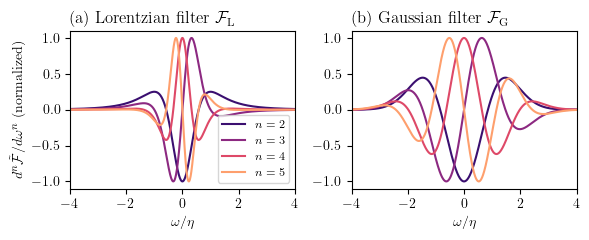}
\end{center}
\vspace{-15pt}
\caption{$n$-th order derivative of the filter, $d^n\tilde{\mathcal{F}}/d\omega^n$ ($n=2,3,4,$ $5$), as a function of frequency $\omega$. Here, $\mathcal{F}\in\{\mathcal{F}_{\rm L}, \mathcal{F}_{\rm G}\}$, and all curves are normalized for comparison.}
\label{fig_7_filter_deriv}
\end{figure}

In this appendix, we derive the subleading term in Eq.~\eqref{noisy_spectral_function_depolarizing} from Eq.~\eqref{noisy_density_matrix_depolarizing}. We simplify Eq.~\eqref{noisy_density_matrix_depolarizing} by rearranging $U_D\cdots U_2$ $U_1 = [U_{\rm I}(\vec{\theta})]^\dag U_M(t)$ $U_{\rm I}(\vec{\theta})$ and using $\ket{\psi_{\rm I}} = U_{\rm I}(\vec{\theta}) \prod_{j=0}^{N-1}$ $|0\rangle_{j}^z$ for the specific setup in Fig.~\ref{fig_2_flowchart}(b). Plugging the simplified form in Eq.~\eqref{propagator_noiseless}, the time propagator takes the noisy form:

\begin{equation}
\tilde{\mathcal{P}}_{\rm gd}
= {\rm Tr}[\rho_0\tilde{\rho}_D ]
= (1 - p_{\rm gd}) |\langle \psi_{\rm I}|U_M(t)|\psi_{\rm I}\rangle|^2 + \frac{p_{\rm gd}}{d}.
\label{noisy_time_propagator_depol}
\end{equation}
Notably, in Eq.~\eqref{noisy_time_propagator_depol}, Trotterization is entirely independent of the correction due to depolarizing noise. Thus, we can take the same approach as the noiseless case to expand the Trotterized unitary~\cite{Childs2021}: 
\begin{equation}
U_M (t) = e^{-iHt} + \int_0^t d\tau e^{-iH(t-\tau)} \mathcal{R}(\tau),
\label{unitary_trotter}
\end{equation}
where $\mathcal{R}(t)$ represents the remainder in the expansion: 
\begin{equation}
\mathcal{R}(t) = -\frac{t}{M} \left[H_1,H_2\right] + \mathcal{O}(t^2/M^2).
\label{eq:Rdepol}
\end{equation}
Using the eigenstate expansion $|\psi_{\rm I}\rangle = \sum_u c_u|u\rangle$ with $|u\rangle$ satisfying $H|u\rangle = \mathcal{E}_u|u\rangle$, we find the expansion:
\begin{align}
& \langle \psi_{\rm I}|U_M(t)|\psi_{\rm I}\rangle
= \sum_u |c_u|^2 e^{-i\mathcal{E}_u t} 
\nonumber\\
& ~~~~~~~ + M^{-1} \sum_u \sum_{n=2}^\infty \tilde{c}_{un}^* c_u (-iht)^n + \mathcal{O}(M^{-2}),
\label{time_propagator_nsq}
\\
& |\langle \psi_{\rm I}|U_M(t)|\psi_{\rm I}\rangle|^2 
= \sum_{u,u'} |c_u|^2 |c_{u'}|^2 e^{-i(\mathcal{E}_u - \mathcal{E}_{u'})t} 
\nonumber\\
& ~~~~~~~ + 2 M^{-1} \sum_{u,u'} \sum_{n=2}^\infty |c_u|^2 {\rm Re}\big[ \tilde{c}_{u'n} c_{u'}^* (iht)^n e^{-i\mathcal{E}_u t} \big]
\nonumber\\
& ~~~~~~~ + \mathcal{O}(M^{-2}),
\label{time_propagator_sq}
\end{align}
where we define $\tilde{c}_{un}^* = \frac{1}{n!}h^{-n} \sum_v c_v^* \mathcal{E}_v^{n-2} \bra{v} [H_1,H_2]\ket{u}$, with the energy unit $h\in\{h_1,h_2\}$ chosen for normalization. In the derivation, we used the expansion $e^x = \sum_{n=0}^\infty x^n/n!$, where the $n=0,1$ terms cancel due to other contributions. 

Following the same approach as in deriving Eq.~\eqref{noiseless_spectral_function_Lehmann}, we substitute Eqs.~\eqref{noisy_time_propagator_depol} and \eqref{time_propagator_sq} into Eqs.~\eqref{spectral_function_convolution}-\eqref{spectral_function_continuum} to derive the noisy spectral function, as given in Eq.~\eqref{noisy_spectral_function_depolarizing}. In the noiseless part, the leading term described by Eq.~\eqref{noiseless_spectral_function_Lehmann} is accompanied by the subleading term defined as:
\begin{equation}
\delta\mathcal{A}(\omega) = \sum_u \sum_{n=2}^\infty w_{un} \sum_{s=\pm} \frac{d^n\tilde{\mathcal{F}}(\omega-s\mathcal{E}_u)}{d(s\omega/h)^n},
\label{spectral_function_Trotter_error_noiseless}
\end{equation}
along with the spectral weight $w_{un} = |c_u|^2\sum_{u'} {\rm Re}[\tilde{c}_{u'n}^*c_{u'}]$. In the derivation, we used the integral formula: $\int dt\,t^n e^{i\omega t}$ $= 2\pi i^n d^n \delta(\omega)/d\omega^n$, where $\delta(x)$ is the Dirac delta function. Notably, while the leading term [Eq.~\eqref{noiseless_spectral_function_Lehmann}] is positive definite, the subleading term [Eq.~\eqref{spectral_function_Trotter_error_noiseless}] is not, as both $w_{un}$ and $d^n\tilde{\mathcal{F}}/d\omega^n$ ($n\geq 2$) generally fail to meet this criterion (see Fig.~\ref{fig_7_filter_deriv}). The line shape for $n=2$ was used to schematize the satellite peaks in Fig.~\ref{fig_1_schematic}.

\section{Propagation of Pauli strings on noisy circuits}
\label{appendix_Pauli_propagation}

\begin{figure}[b]
\begin{center}
\includegraphics[width=0.42\textwidth]{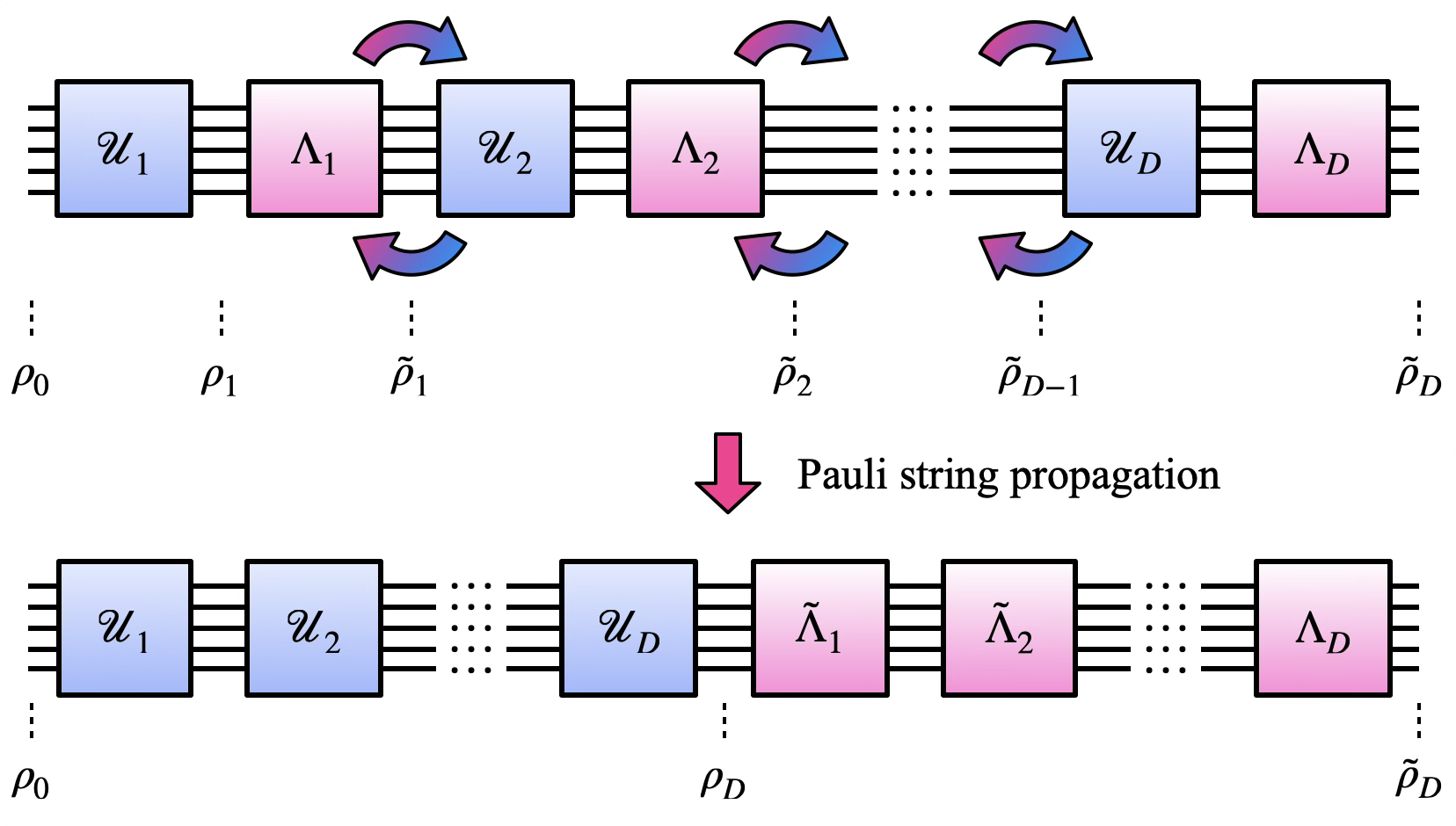}
\end{center}
\vspace{-15pt}
\caption{Construction of an equivalent noisy quantum circuit to the original [Fig.~\ref{fig_3_block_diagram_noise}(b)] under general Markovian noise channels. Using Pauli string propagation, we successively swap the unitary maps $\mathcal{U}_\nu$ and the noise maps $\Lambda_{\nu'}$ for $\nu,\nu'\in[1,D]$. All $\Lambda_\nu$'s are pushed to the circuit end to define the modified noise maps $\tilde{\Lambda}_\nu$'s acting on $\rho_0$.}
\label{fig_8_branchingdiagram_nondepolarizing}
\end{figure}

In the derivation presented in Sec.~\ref{sec_Markovian}, the propagation of Pauli strings plays a crucial role in swapping the unitary maps $\mathcal{U}_{\nu\in[2,D]}$ and the mid-circuit noise maps $\Lambda_{\nu\in[1,D-1]}$, effectively shifting all noise maps to the circuit end (see Fig.~\ref{fig_8_branchingdiagram_nondepolarizing}). The result is formulated as follows:
\begin{gather}
\tilde{\rho}_D = (\Lambda_D\circ\tilde{\Lambda}_{D-1}\circ\cdots\circ\tilde{\Lambda}_2\circ\tilde{\Lambda}_1)(\rho_D),
\label{noisy_density_matrix_generalized_pauli}
\\
\rho_D = (\mathcal{U}_D\circ\cdots\circ\mathcal{U}_2\circ\mathcal{U}_1)(\rho_0),
\label{noisy_input_density_matrix_generalized_pauli}
\end{gather}
where the noise map $\tilde{\Lambda}_\nu$ is modified from Eq.~\eqref{Generalized_Pauli_map}. At the operator level, a similarity transform is employed to give a rule for swapping the unitary $U_\nu$ and the Kraus operator $K_{k_\nu q_\nu}^{(\nu)}$ [in Eq.~\eqref{Generalized_Pauli_map_branches}], which generally do not commute.  

Pauli string propagation is exemplified by the TFIM circuit. Here, without loss of generality, we set $U_\nu \in \{R^x,R^{zz}\}$ and represent $K_{k_\nu q_\nu}^{(\nu)}$ by Pauli strings. For $U_\nu = R^x$, the similarity transform is nothing but the Rodrigues' rotation formula:
\begin{align}
& ~R_j^x\left(\frac{2Jt}{M}\right) 
\left\{
\begin{array}{c}
\sigma_j^y \\
\sigma_j^z
\end{array}
\right\}
R_j^x\left(\frac{-2Jt}{M}\right)
\nonumber\\
& = \left\{
\begin{array}{c}
\sigma_j^y 
\\
\sigma_j^z 
\end{array}
\right\} \cos\left(\frac{2Jt}{M}\right) + \left\{
\begin{array}{c}
\sigma_j^z 
\\
- \sigma_j^y
\end{array}
\right\} \sin\left(\frac{2Jt}{M}\right).
\label{similarity_transform_1}
\end{align}
This formula can be extended to deal with the case $U_\nu = R^{zz}$. Some notable results include:
\begin{align}
& ~R_{j,j+1}^{zz}\left(\frac{2Jt}{M}\right) 
\left\{
\begin{array}{c}
\sigma_j^x \sigma_{j+1}^z \\
\sigma_j^y \sigma_{j+1}^z \\
\sigma_j^z \sigma_{j+1}^x \\
\sigma_j^z \sigma_{j+1}^y
\end{array}
\right\}
R_{j,j+1}^{zz}\left(\frac{-2Jt}{M}\right) 
\nonumber\\
& = \left\{
\begin{array}{c}
\sigma_j^x\sigma_{j+1}^z
\\
\sigma_j^y\sigma_{j+1}^z
\\
\sigma_j^z\sigma_{j+1}^x
\\
\sigma_j^z\sigma_{j+1}^y
\end{array}
\right\} \cos\left(\frac{2Jt}{M}\right) + \left\{
\begin{array}{c}
\sigma_j^y \sigma_{j+1}^0
\\
- \sigma_j^x \sigma_{j+1}^0
\\
\sigma_j^0 \sigma_{j+1}^y
\\
- \sigma_j^0 \sigma_{j+1}^x
\end{array}
\right\} \sin\left(\frac{2Jt}{M}\right),
\label{similarity_transform_2}
\end{align}
where $\sigma^0 \equiv \mathbb{I}_2$. Note that, in Eqs.~\eqref{similarity_transform_1} and \eqref{similarity_transform_2}, an additional time dependency is introduced by the propagation of Pauli strings over the unitary.

\section{Derivation of the subleading term in Eq.~\eqref{noisy_spectral_function_nondepolarizing}}
\label{appendix_derivation_nondepolarizing}

In this appendix, we derive the subleading term in Eq.~\eqref{noisy_spectral_function_nondepolarizing} from Eqs.~\eqref{noisy_density_matrix_generalized_pauli}-\eqref{noisy_input_density_matrix_generalized_pauli}. We proceed in a similar manner to Appendix~\ref{appendix_derivation_depolarizing}. In this case, the time propagator takes the noisy form: 
\begin{equation}
\tilde{\mathcal{P}}_{\rm gp} = {\rm Tr}[\rho_0\tilde{\rho}_D] = p_{\vec{0}} |\langle \psi_{\rm I}|U_M(t)|\psi_{\rm I}\rangle|^2 + \sum_{\vec{k}\neq\vec{0}} p_{\vec{k}} \mathcal{P}_{\vec{k}},
\label{noisy_time_propagator_nondepol}
\end{equation}
where the correction due to the noise channel $\vec{k}$ is defined as:
\begin{align}
\mathcal{P}_{\vec{k}} 
& = \frac{1}{(d-1)^D} \sum_{\vec{q}} |\mathcal{Q}_{\vec{k}\vec{q}}|^2,
\label{propagator_noisy_nondepol}
\\
\mathcal{Q}_{\vec{k}\vec{q}} 
& = \langle \psi_{\rm I}|U_{\rm I}(\vec{\theta})K_{k_Dq_D}^{(D)}\tilde{K}_{k_{D-1} q_{D-1}}^{(D-1)} \cdots\tilde{K}_{k_2 q_2}^{(2)}\tilde{K}_{k_1 q_1}^{(1)}
\nonumber\\
& ~\times [U_{\rm I}(\vec{\theta})]^\dag U_M(t) |\psi_{\rm I}\rangle.
\label{propagator_nonsq_nondepol}
\end{align}
In contrast to the last term in Eq.~\eqref{noisy_time_propagator_depol}, Eqs.~\eqref{propagator_noisy_nondepol}-\eqref{propagator_nonsq_nondepol} are not simply addressed mainly due to additional time dependency in $\tilde{K}_{k_\nu q_\nu}^{(\nu)}$. 
In general, $\tilde{K}_{k_\nu q_\nu}^{(\nu)}$ is time-periodic with two distinct periods, $\pi M/|h_1|$ and $\pi M/|h_2|$, where $h_{1,2}$ are the coefficients of Pauli strings in $H_{1,2}$. Thus we can make double series expansion:
\begin{align}
& \tilde{K}_{k_{D-1} q_{D-1}}^{(D-1)}(t) \cdots\tilde{K}_{k_2 q_2}^{(2)}(t) \tilde{K}_{k_1 q_1}^{(1)}(t) 
\nonumber\\
& = \sum_{r_1,r_2} \mathbb{K}_{k_{D-1}\cdots k_2 k_1;q_{D-1}\cdots q_2 q_1}^{(r_1r_2)} e^{-2i(r_1h_1 + r_2h_2)t/M},
\label{double_series_expansion}
\end{align}
where the expansion coefficient $\mathbb{K}_{k_{D-1}\cdots k_2 k_1;q_{D-1}\cdots q_2 q_1}^{(r_1r_2)}$ informs on the details of noise channels, and both $r_1$, $r_2$ are supposed to have dependency on $\vec{k}$, $\vec{q}$. To proceed further, we need to deal with $U_M(t)$, given by Eq.~\eqref{unitary_trotter}. Substituting into Eq.~\eqref{propagator_nonsq_nondepol} along with Eq.~\eqref{double_series_expansion}, we find the expansion: 
\begin{align}
\mathcal{Q}_{\vec{k}\vec{q}} 
& = \sum_{u,r_1,r_2} [\tilde{c}_{ur_1r_2}^{(\vec{k}\vec{q})}]^* c_u e^{-i[\mathcal{E}_u + 2(r_1h_1 + r_2h_2)/M]t}
\nonumber\\
& ~~~ + M^{-1} \sum_{u,r_1,r_2} \sum_{n=2}^\infty [\tilde{d}_{ur_1r_2n}^{(\vec{k}\vec{q})}]^* c_u (-iht)^n 
\nonumber\\
& ~~~ \times e^{-2i(r_1h_1 + r_2h_2)t/M} + \mathcal{O}(M^{-2}),
\label{time_propagator_noisy_nonsq}
\end{align}
\begin{align}
|\mathcal{Q}_{\vec{k}\vec{q}}|^2
& = \sum_{u,u'} \sum_{r_1,r'_1} \sum_{r_2,r'_2} [\tilde{c}_{ur_1r_2}^{(\vec{k}\vec{q})}]^* c_u c_{u'}^* [\tilde{c}_{u'r'_1r'_2}^{(\vec{k}\vec{q})}] 
\nonumber\\
& ~~~ \times e^{-i\{\mathcal{E}_u-\mathcal{E}_{u'} + 2M^{-1}[(r_1-r'_1)h_1 + (r_2-r'_2)h_2]\}t}
\nonumber\\
& ~~~ + 2M^{-1} \sum_{u,u'} \sum_{r_1,r'_1} \sum_{r_2,r'_2} \sum_{n=2}^\infty  
\nonumber\\
& ~~~ \times {\rm Re} \big\{ [\tilde{c}_{ur_1r_2}^{(\vec{k}\vec{q})}]^* c_u c_{u'}^* [\tilde{d}_{u'r'_1r'_2n}^{(\vec{k}\vec{q})}] (iht)^n 
\nonumber\\
& ~~~ \times e^{-i\{\mathcal{E}_u + 2M^{-1}[(r_1-r'_1)h_1 + (r_2-r'_2)h_2]\}t} \big\}
\nonumber\\
& ~~~  + \mathcal{O}(M^{-2}),
\label{time_propagator_noisy_sq}
\end{align}
where the noisy expansion coefficients are defined as:
\begin{align}
[\tilde{c}_{ur_1r_2}^{(\vec{k}\vec{q})}]^* 
& = \sum_{u'} c_{u'}^* \langle u'|U_{\rm I}(\vec{\theta})K_{k_Dq_D}^{(D)} 
\nonumber\\
& ~\times \mathbb{K}_{k_{D-1}\cdots k_2 k_1;q_{D-1}\cdots q_2 q_1}^{(r_1r_2)} [U_{\rm I}(\vec{\theta})]^\dag |u\rangle,
\label{spectral_weigh_noisy_c}
\\
[\tilde{d}_{ur_1r_2n}^{(\vec{k}\vec{q})}]^* 
& = \frac{h^{-n}}{n!} \sum_v [\tilde{c}_{vr_1r_2}^{(\vec{k}\vec{q})}]^* \mathcal{E}_v^{n-2} \bra{v}[H_1,H_2]\ket{u},
\label{spectral_weigh_noisy_d}
\end{align}
with the energy unit $h\in\{h_1,h_2\}$ chosen for normalization.

The remaining task proceeds in the same manner as Appendix~\ref{appendix_derivation_depolarizing}. Plugging Eqs.~\eqref{noisy_time_propagator_nondepol}, \eqref{propagator_noisy_nondepol} and \eqref{time_propagator_noisy_sq} in Eqs.~\eqref{spectral_function_convolution}-\eqref{spectral_function_continuum}, we can derive the noisy spectral function, Eq.~\eqref{noisy_spectral_function_nondepolarizing}. The noiseless part ($\vec{k}=\vec{0}$) is given by Eqs.~\eqref{noiseless_spectral_function_Lehmann} and \eqref{spectral_function_Trotter_error_noiseless}. In the noisy part ($\vec{k}\neq\vec{0}$), the leading term described by Eq.~\eqref{spectral_function_noisy_correction} is accompanied by the subleading term defined as:
\begin{align}
\delta\tilde{\mathcal{A}}_{\rm gp}^{(\vec{k})}(\omega) 
& = \sum_u \sum_{r_1,r'_1} \sum_{r_2,r'_2} \sum_{n=2}^\infty \tilde{w}_{ur_1r'_1r_2r'_2n}^{(\vec{k})} 
\nonumber\\
& ~\times \sum_{s=\pm} \frac{d^n\tilde{\mathcal{F}}(\omega-s\mathcal{E}_u-s\delta\Delta_{r_1r'_1r_2r'_2})}{d(s\omega/h)^n},
\label{spectral_function_Trotter_error_noisy}
\end{align}
where $\delta\Delta_{r_1r'_1r_2r'_2}$ is given by Eq.~\eqref{gap_nondepol}, and the associated spectral weight is modified from Eq.~\eqref{spectral_weight_nondepol}:
\begin{equation}
\tilde{w}_{ur_1r'_1r_2r'_2n}^{(\vec{k})} = \frac{1}{(d-1)^D} \sum_{\vec{q}} \sum_{u'} {\rm Re} \big\{ [\tilde{c}_{ur_1r_2}^{(\vec{k}\vec{q})}]^* c_u c^*_{u'} \tilde{d}_{u'r'_1r'_2n}^{(\vec{k}\vec{q})} \big\}.
\end{equation}
Notably, Eq.~\eqref{spectral_function_Trotter_error_noisy}, similar to Eq.~\eqref{spectral_function_Trotter_error_noiseless}, lacks positive definiteness.

\section{Baseline correction with asymmetric least squares}
\label{appendix_ALS}

In this paper, ALS fitting is crucial for baseline correction in noisy spectral data, enhancing the accuracy of gap estimation. This section provides a detailed explanation of the ALS fitting process~\cite{Eilers2005}. See Ref.~\cite{Zhang2010, He2014, Baek2015, Korepanov2020} for additional variations.

We consider a discrete series of the original spectral function, $\{\mathcal{A}_m\}_{m=0}^{L-1}$, where $\mathcal{A}_m = \mathcal{A}(\omega_m)$. Our goal is to determine the smoothing series $\{\mathcal{B}_m\}_{m=0}^{L-1}$, which effectively captures the baseline of the original series. This is achieved by minimizing the penalized least squares function:
\begin{equation}
f_{\rm ALS} = \sum_m w_m (\mathcal{A}_m - \mathcal{B}_m)^2 + \lambda \sum_m (D \mathcal{B}_m)^2.
\label{least_squares_function}
\end{equation}
The first term in Eq.~\eqref{least_squares_function} fits the series with weight:
\begin{align}
w_m = \left\{
\begin{array}{cc}
\chi, & \mathcal{A}_m > \mathcal{B}_m
\\
1 - \chi, & \mathcal{A}_m < \mathcal{B}_m
\end{array}
\right.
\label{weights_als}
\end{align}
where $\chi$ is the asymmetric parameter, typically ranging from $10^{-3}$ to $10^{-1}$. 
This range effectively suppresses negative deviations while maintaining flexibility in fitting the baseline beneath the signal peaks. The second term acts as a penalty for non-smooth behavior in $\mathcal{B}$. Here, $D$ represents the second-order difference: $D \mathcal{B}_m = (\mathcal{B}_m - \mathcal{B}_{m-1}) - (\mathcal{B}_{m-1} - \mathcal{B}_{m-2}) = \mathcal{B}_m - 2\mathcal{B}_{m-1} + \mathcal{B}_{m-2}$, and $\lambda$ is the smoothing parameter that adjusts the balance between two terms, typically ranging from $10^2$ to $10^9$, with flexibility to deviate from this range depending on data characteristics and desired baseline smoothness.

The minimization problem in Eq.~\eqref{least_squares_function} simplifies to finding a solution for the system of equations: 
\begin{equation}
(\mathbb{W} + \lambda \mathbb{D}^T \mathbb{D}) \vec{\mathcal{B}} = \mathbb{W} \vec{\mathcal{A}},
\label{solution_als}
\end{equation}
where we define the matrices $\mathbb{W} = {\rm diag}(w_0,w_1,\cdots,w_{L-1})$ and $\mathbb{D}$ such that $\mathbb{D}\vec{\mathcal{B}} = D\vec{\mathcal{B}}$ for the vector $\vec{\mathcal{B}} = (\mathcal{B}_0,\mathcal{B}_1,\cdots,$ $\mathcal{B}_{L-1})$. As Eq.~\eqref{solution_als} interacts with Eq.~\eqref{weights_als}, the solution can be obtained iteratively, for example, beginning with uniform weights as the initial condition. Typically, a few dozen iterations are sufficient for convergence.

\section{Qiskit code overview}
\label{appendix_qiskit_overview}

In this appendix, we outline the code used in the Qiskit implementation of the hybrid QGE algorithm. The complete Qiskit code can be assessed~\cite{GitHub}.

\vspace{5mm}

{\centering ALGORITHM 1. Pseudocode for the hybrid QGE algorithm \par}

\noindent\rule{0.49\textwidth}{0.4pt}\\
[-9pt]
\noindent\rule{0.49\textwidth}{0.4pt}
\\
{\bf Initial setup}
\vspace{-2mm}
\\
\noindent\rule{0.49\textwidth}{0.4pt}\\
[-9pt]
\noindent\rule{0.49\textwidth}{0.4pt}
\begin{algorithmic}[1]
\State Quantum process: IBM Quantum backend, qubit number ($N$) and initial layout, coupling ratio ($J/h$), Trotter circuit depth ($M$), and measurement shots
\State Postprocess: filter ($\mathcal{F}$), broadening ($\eta$), Fourier sampling number ($L$), and parameters for fitting and optimization
\State Estimate the initial gap $\Delta_0$ using perturbation theory.
\State Evaluate the exact gap $\Delta_{\rm exact}$ via exact diagonalization.
\end{algorithmic}
\vspace{-2mm}
\noindent\rule{0.49\textwidth}{0.4pt}\\
[-9pt]
\noindent\rule{0.49\textwidth}{0.4pt}
\\
{\bf Main routine} \Comment{Execute both Fourier sampling and trial-state optimization in session mode.}
\vspace{-2mm}
\\
\noindent\rule{0.49\textwidth}{0.4pt}\\
[-9pt]
\noindent\rule{0.49\textwidth}{0.4pt}
\begin{algorithmic}[1]
\State {\bf with} \textsc{session}(backend=backend) {\bf as} session
\State \hspace{4mm} Call \textsc{objective}($\theta$, session) to evaluate $f(\theta)$.
\State \hspace{4mm} Call \textsc{minimize\_scalar}($f(\theta)$, method=`bounded',
\State \hspace{4mm} bounds=$[0,\pi/2]$, tolerance=$10^{-6}$, args=(session,))
\end{algorithmic}
\vspace{-2mm}
\noindent\rule{0.49\textwidth}{0.4pt}\\
[-9pt]
\noindent\rule{0.49\textwidth}{0.4pt}
\\
{\bf Subroutine:} \textsc{objective}($\theta$, session)
\vspace{-2mm}
\\
\noindent\rule{0.49\textwidth}{0.4pt}\\
[-9pt]
\noindent\rule{0.49\textwidth}{0.4pt}
\begin{algorithmic}[1]
\For{$s = \pm 1$} \Comment{$s=1(-1)$: (anti-)causal}
    \For{$n = 0$ to $L-1$} \Comment{Fourier sampling}
        \State Call \textsc{sampler}(mode=session). 
        \State Call \textsc{circuits}($\theta$, $n$, $s$) to build quantum circuits.
        \State Run quantum circuits in session mode.
        \State Evaluate the filtered time propagator $\mathcal{F}_n \mathcal{P}_{sn}$.
    \EndFor
    \State Perform the fast Fourier transform for $\{\mathcal{F}_n \mathcal{P}_{sn}\}_{n=0}^{L-1}$.
\EndFor
\State Evaluate the spectral function $\{\mathcal{A}(\omega_m)\}_{m=0}^{L-1}$ [Eq.~\eqref{spectral_function}].
\State Set the target window $\omega_m\in[\Delta_0-\eta,\Delta_0+\eta]$.
\State Search for the bare gap $\Delta_{\rm bare}$ locating the target signal peak nearest to $\Delta_0$.
\State Evaluate the error $|\Delta_{\rm bare}-\Delta_{\rm exact}|/\Delta_{\rm exact}$.
\State Call \textsc{asls}($\{\mathcal{A}(\omega_m)\}_{m=0}^{L-1}$, $\lambda=1$, $\chi=10^{-2}$) to evaluate the baseline correction to $\mathcal{A}$ using ALS fitting.
\State Set the target window $\omega_m\in[\Delta_0-\eta,\Delta_0+\eta]$.
\State Search for the corrected gap $\Delta_{\rm corr}$ locating the target signal peak nearest to $\Delta_0$.
\State Evaluate the error $|\Delta_{\rm corr}-\Delta_{\rm exact}|/\Delta_{\rm exact}$.
\State Evaluate the cost function $f(\theta) = [h\mathcal{A}(\omega_m=\Delta_{\rm corr})]^{-1}$.
\State Store results to disk.
\State \Return $f(\theta)$
\end{algorithmic}
\vspace{-2mm}
\noindent\rule{0.49\textwidth}{0.4pt}\\
[-9pt]
\noindent\rule{0.49\textwidth}{0.4pt}
\\
{\bf Subroutine:} \textsc{circuits}($\theta$, $n$, $s$)
\vspace{-2mm}
\\
\noindent\rule{0.49\textwidth}{0.4pt}\\
[-9pt]
\noindent\rule{0.49\textwidth}{0.4pt}
\begin{algorithmic}[1]
\State Initialize quantum and classical registers.
\State Apply the trial-state unitary $U_{\rm I}(\theta)$ [Eq.~\eqref{trial_state_unitary}].
\State Apply the Trotterized unitary $U_M(t_{sn})$ [Eq.~\eqref{Trotterized_unitary}].
\State Apply the inverse trial-state unitary $[U_{\rm I}(\theta)]^\dag$.
\State Measure the output state on the same circuit.
\State \Return measurement outcomes
\end{algorithmic}
\vspace{-2mm}
\noindent\rule{0.49\textwidth}{0.4pt}\\
[-9pt]
\noindent\rule{0.49\textwidth}{0.4pt}
\\
{\bf Subroutine:} \textsc{asls}($\{\mathcal{A}(\omega_m)\}_{m=0}^{L-1}$, $\lambda$, $p$)
\vspace{-2mm}
\\
\noindent\rule{0.49\textwidth}{0.4pt}\\
[-9pt]
\noindent\rule{0.49\textwidth}{0.4pt}
\begin{algorithmic}[1]
\State Set the initial weight $\vec{w}_0$.
\For{$i = 1$ to $i_{\rm max}$} \Comment{Iteration}
    \State Construct the matrix $\mathbb{W} + \lambda \mathbb{D}^T\mathbb{D}$.
    \State Find $\{\mathcal{B}(\omega_m)\}_{m=0}^{L-1}$ by solving Eq.~\eqref{solution_als}.
    \State Update the weight $\vec{w}_i$ using Eq.~\eqref{weights_als}.
    \If{$|\vec{w}_i-\vec{w}_{i-1}| <$ tolerance}
        \State Break
    \EndIf
\EndFor
\State \Return baseline function $\{\mathcal{B}(\omega_m)\}_{m=0}^{L-1}$
\end{algorithmic}
\vspace{-2mm}
\noindent\rule{0.49\textwidth}{0.4pt}\\
[-9pt]
\noindent\rule{0.49\textwidth}{0.4pt}
\\
{\bf Subroutine:} \textsc{minimize\_scalar}($f(\theta)$, method=`bounded', \\bounds, tolerance, args)
\vspace{-2mm}
\\
\noindent\rule{0.49\textwidth}{0.4pt}\\
[-9pt]
\noindent\rule{0.49\textwidth}{0.4pt}
\begin{algorithmic}[1]
\For{$i = 1$ to $i_{\rm max}$} \Comment{Iteration}
    \State Evaluate $f$ at the midpoint $\theta_i$ of the current bounds.
    \If{$i\geq 2$ and $|f(\theta_i) - f(\theta_{i-1})| <$ tolerance}
        \State Break
    \EndIf
    \State Evaluate $f$ at two endpoints of the current bounds.
    \State Select the sub-interval that returns the lower $f$.
    \State Update the bounds to reflect the selected sub-interval.
\EndFor
\State \Return optimal value of $\theta$ that minimizes $f$
\end{algorithmic}
\vspace{-2mm}
\noindent\rule{0.49\textwidth}{0.4pt}\\
[-9pt]
\noindent\rule{0.49\textwidth}{0.4pt}

\clearpage

\section{IBM Quantum backend calibration data}
\label{appendix_backend_calibration}

\begin{figure}[h]
\begin{center}
\includegraphics[width=0.45\textwidth]{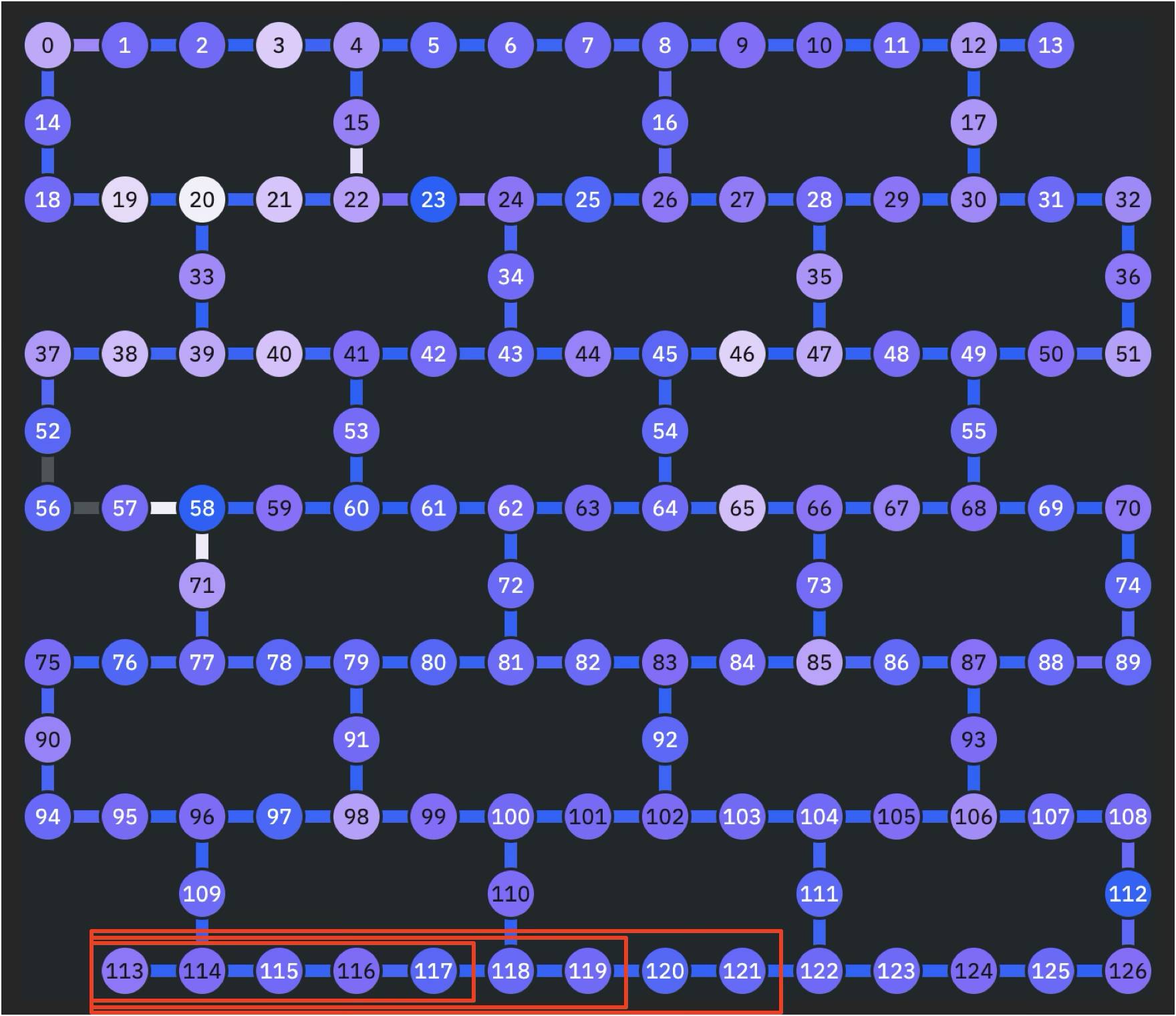}
\end{center}
\vspace{-15pt}
\caption{Qubit layout for the {\it ibm\_sherbrooke} processor. For our simulations, we used linear qubit arrays (highlighted by red boxes): 113–117 (5 qubits), 113–119 (7 qubits), and 113–121 (9 qubits). The color map for each qubit represents, e.g., $T_1$, where darker shades correspond to the lower bound and lighter shades to the upper bound.}
\label{fig_9_ibm_sherbrooke}
\end{figure}

In this appendix, we present details of the {\it ibm\_sherbrooke} processor, which was used for both noisy simulations and real-device demonstrations. The qubit layout is illustrated in Fig.~\ref{fig_9_ibm_sherbrooke}, and calibration data is provided in Table~\ref{table_device_calibrations}.

\vspace{-5mm}

\begin{table}[h]
\centering
\footnotesize
\begin{tabular}{|c|c|c|c|c|c|c|c|c|c|}
\hline
(a) & (b) & (c) & (d) & (e) & (f) & (g) & (h) & (i) & (j) \\
\hline\hline
{\bf 113} & 315.7 & 69.70  & 4.963 & -0.30844 & 5.6 & 6.2 & 5.0 & 2.080 & \multirow{2}{*}{{\bf 113-114}:5.618} \\
\cline{1-9}
{\bf 114} & 273.6 & 383.3 & 4.885 & -0.30934 & 13.6 & 13.0 & 14.2 & 1.598 & \multirow{2}{*}{{\bf 114-115}:8.790} \\
\cline{1-9}
{\bf 115} & 236.7 & 158.4 & 4.743 & -0.31098 & 4.6 & 6.0 & 3.2 & 1.956 & \multirow{2}{*}{{\bf 115-116}:7.296} \\
\cline{1-9}
{\bf 116} & 275.5 & 384.7 & 4.931 & -0.30852 & 9.2 & 9.0 & 9.4 & 1.777 & \multirow{2}{*}{{\bf 116-117}:12.85} \\
\cline{1-9}
{\bf 117} & 163.8 & 193.2 & 4.793 & -0.31084 & 8.4 & 9.8 & 7.0 & 6.601 & \multirow{2}{*}{{\bf 117-118}:5.905} \\
\cline{1-9}
{\bf 118} & 208.0 & 227.9 & 4.737 & -0.31214 & 5.8 & 8.0 & 3.6 & 1.523 & \multirow{2}{*}{{\bf 118-119}:7.019} \\
\cline{1-9}
{\bf 119} & 234.4 & 178.4 & 4.793 & -0.31150 & 5.7 & 7.4 & 4.0 & 6.059 & \multirow{2}{*}{{\bf 119-120}:7.777} \\
\cline{1-9}
{\bf 120} & 147.9 & 180.3 & 4.888 & -0.30955 & 8.1 & 6.8 & 9.4 & 1.981 & \multirow{2}{*}{{\bf 120-121}:5.819} \\
\cline{1-9}
{\bf 121} & 199.7 & 21.18  & 4.848 & -0.31085 & 10.8 & 9.2 & 12.4 & 3.815 & \\
\hline
\end{tabular}
\caption{Calibration data for the {\it ibm\_sherbrooke} processor (collected on 10/23/2024). Each column provides data for: (a) qubit, (b) $T_1$ ($\mu$s), (c) $T_2$ ($\mu$s), (d) frequency (GHz), (e) anharmonicity (GHz), (f) readout assignment error ($\times 10^{-3}$), (g) probability of $\ket{0}$ measurement for $\ket{1}$ preparation ($\times 10^{-3}$), (h) probability of $\ket{1}$ measurement for $\ket{0}$ preparation ($\times 10^{-3}$), (i) identity, $\sqrt{X}$, or Pauli-$X$ error ($\times 10^{-4}$), (j) ECR error ($\times 10^{-3}$). Shared parameters across all qubits include: readout length = 1216 ns, gate time = 533.3 ns, and $R^z$ error = 0.}
\label{table_device_calibrations}
\end{table}

\vspace{-5mm}

\section{Supplementary simulation data for Figs.~\ref{fig_5_optimization_N5_FL}-\ref{fig_6_convergence_FL_compact}}
\label{appendix_QGE_procedure}

\begin{figure*}[t]
\begin{center}
\includegraphics[width=0.96\textwidth]{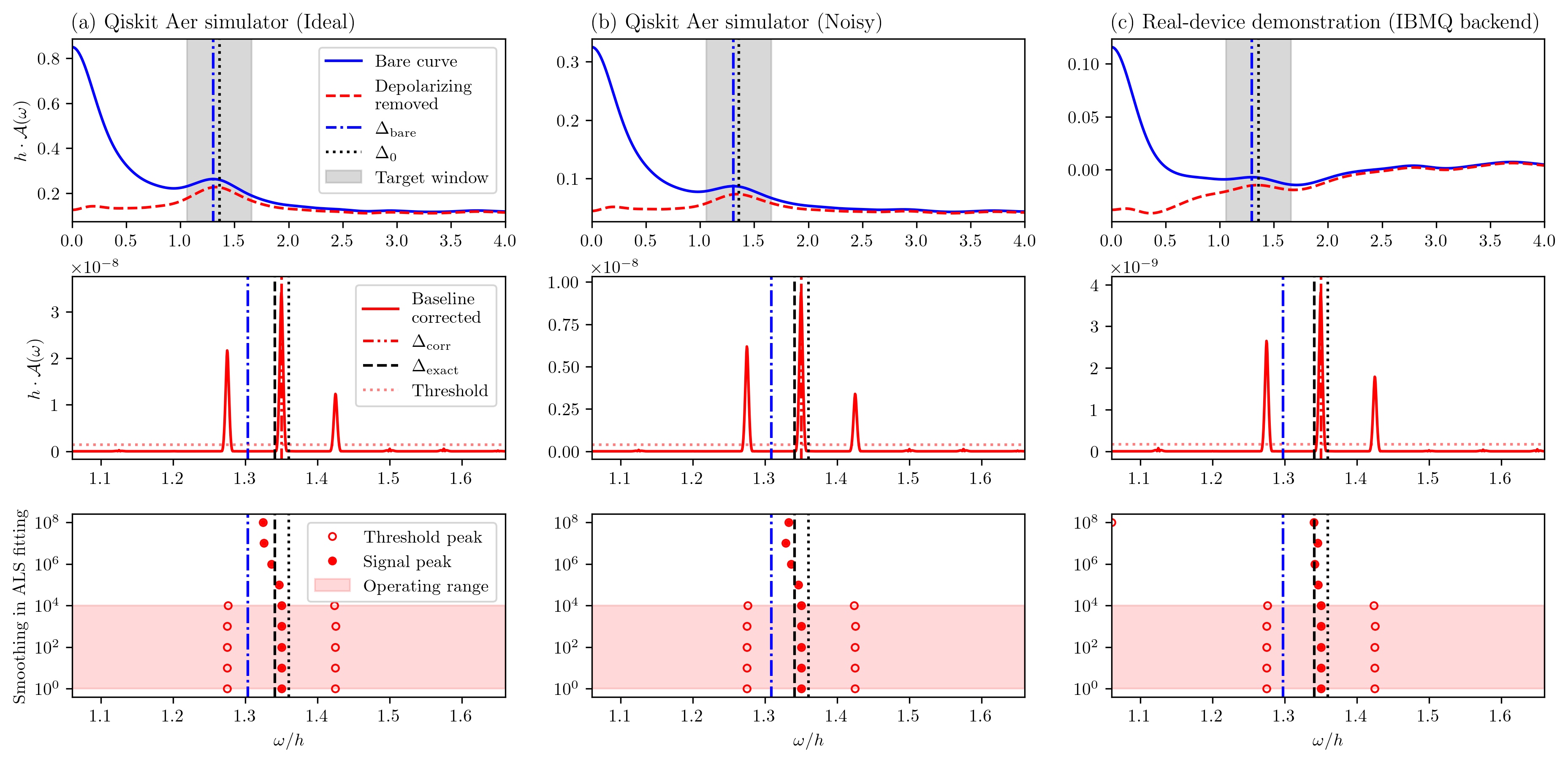}
\end{center}
\vspace{-15pt}
\caption{Simulation results illustrating the QGE procedure using different approaches corresponding to Fig.~\ref{fig_5_optimization_N5_FL}(a)-(c), respectively. Top panels: Bare spectral function $\mathcal{A}$ (blue solid curve) versus frequency $\omega$ for the TFIM of $N=5$, $J/h=0.4$, and the filter $\mathcal{F}=\mathcal{F}_{\rm L}$ with the broadening $\eta/h=0.3$. The input orientation $\theta$ is conveniently fixed to the value obtained at the intermediate step of trial-state optimization (see Fig.~\ref{fig_5_optimization_N5_FL}). The red dashed curve represents 
the adjusted line shape with the redundant peak at $\omega=0$ removed by fitting. The blue dot-dashed vertical line indicates the bare gap estimate $\Delta_{\rm bare}$, which corresponds to the closest peak to the initial gap $\Delta_0$ (black dotted vertical line) in the original data. The target window, defined as $\Delta_0-\eta\leq\omega\leq\Delta_0+\eta$ (shaded in gray), is used to apply the ALS fitting method for baseline correction. Middle panels: Baseline-corrected spectral function (red solid curve). For ALS fitting, the asymmetric and smoothing parameters are set to $\chi = 10^{-2}$ and $\lambda = 1$, respectively. Corrected gap estimate $\Delta_{\rm corr}$ (red double-dot-dashed vertical line) is obtained by estimating the closest peak position to $\Delta_0$ within the target window. The pink dotted horizontal line marks the threshold for peak height. In (a)-(c), $\Delta_{\rm corr}$ is improved over $\Delta_{\rm bare}$ and provides better estimate close to the exact gap $\Delta_{\rm exact}$. Bottom panels: Illustration of the ALS fitting’s operating range (shaded in pink) by adjusting the smoothing parameter. Red open circles indicate the threshold peak positions, while filled circles highlight the target signal peak. Other parameter settings are the same as in Fig.~\ref{fig_5_optimization_N5_FL}.}
\label{fig_10_ALSfitting_N5_FL}
\end{figure*}

\begin{figure*}[t]
\begin{center}
\includegraphics[width=0.92\textwidth]{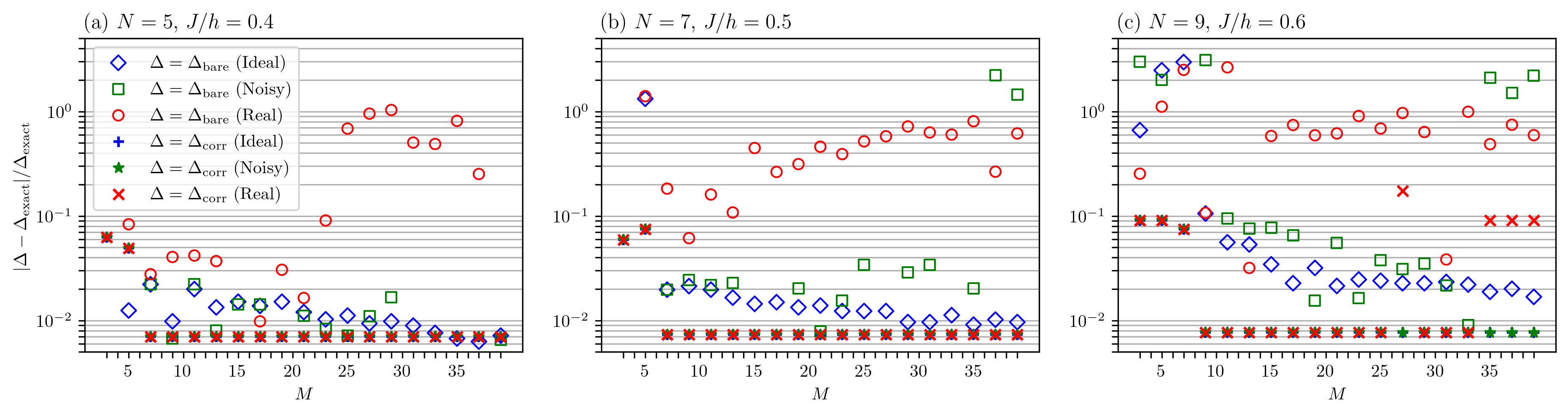}
\end{center}
\vspace{-15pt}
\caption{Extended data corresponding to Fig.~\ref{fig_6_convergence_FL_compact} under the same settings. Each symbol is defined in the legend.}
\label{fig_11_convergence_FL}
\end{figure*}

In this appendix, we provide additional analysis based on the results in Sec.~\ref{sec_numerical_results}. 

(i) {\it QGE procedure}: Figure~\ref{fig_10_ALSfitting_N5_FL} illustrates the QGE procedure with the same parameter settings as in Fig.~\ref{fig_5_optimization_N5_FL}. Here, the input orientation $\theta$ is conveniently set to the value obtained at an intermediate step of trial-state optimization. The top panels show the line shape of the bare spectral function (blue solid curve). As discussed in Sec.~\ref{sec_robustness_to_noise}, the bare gap $\Delta_{\rm bare}$ is estimated by starting with a perturbative solution for a quantum paramagnet, $\Delta_0/h = 2[1-(1-1/N)J/h]$, and identifying the signal peak closest to $\Delta_0$. The largest redundant peak at $\omega = 0$ serves as a primary source of lineshape distortion around the target signal peak, contributing to gap estimate error. As discussed in Sec.~\ref{sec_depolarizing}, it arises from the $u=u'$ term in Eq.~\eqref{noiseless_spectral_function_Lehmann} in the noiseless simulation [panel (a)], and its height is adjusted by a combination of the (scaled) $u=u'$ term and the last term in Eq.~\eqref{noisy_spectral_function_depolarizing} in the simulation dominated by depolarizing noise [panel (b)]. The red dashed curve demonstrates that the redundant peak can be removed from the spectral function through curve fitting. The extent of adjustment varies depending on the degree of line shape spreading. After adjustment, the signal peak is clearly resolved if its height exceeds the device-specific detection threshold, effectively showcasing the algorithm's inherent resilience to depolarizing noise.

The noise resilience argument becomes less robust when other noise types significantly contribute alongside depolarizing noise. Evidence for this appears in the real-device demonstrations in panel (c), where residual background remains even after removing the redundant peak at $\omega=0$. Note that the spectral function becomes negative in the range $\omega\lesssim 2.4h$, indicating a breakdown of positive definiteness. One possible reason is that the noisy spectral weight, Eq.~\eqref{spectral_weight_nondepol}, is not bounded below from zero, unlike the noiseless counterpart $w_{nn'}$ in Eq.~\eqref{noiseless_spectral_function_Lehmann}. As discussed in Secs.~\ref{sec_Markovian} and \ref{sec_robustness_to_noise}, we take an error mitigation strategy based entirely on classical signal processing techniques to effectively remove the residual background and accurately restore the peak location. Specifically, among the available methods, we employ ALS fitting (see Appendix~\ref{appendix_ALS}) to correct the baseline of the spectral function, which has been distorted by the residual background. 

The middle panels in Fig.~\ref{fig_10_ALSfitting_N5_FL} present the baseline-corrected results within the target window $\Delta_0-\eta \leq \omega \leq \Delta_0+\eta$. This correction is achieved through the careful selection of smoothing and asymmetry parameters in the ALS fitting process. The corrected gap $\Delta_{\rm corr}$ is estimated in the same manner as $\Delta_{\rm bare}$. Notably, across all panels (a)-(c), $\Delta_{\rm corr}$ shows an improvement over $\Delta_{\rm bare}$, converging toward $\Delta_{\rm exact}$. The bottom panels illustrate the operating range of the smoothing parameter (below $10^4$) needed to reliably reproduce our results, even with moderate variations in the asymmetry parameter. Here, only the positions of threshold peaks that meet the height threshold ($= \textrm{mean} + k \times \textrm{standard deviation}$; with $k=1$) are shown. This approach minimizes the risk that our findings could be attributed to artifacts of the fitting process.

(ii) {\it Full analysis of scalability and robustness}: We present extended data to complement Fig.~\ref{fig_6_convergence_FL_compact}. While the algorithm's robustness has been evaluated using baseline-corrected spectral data, here we focus on the outcomes for $\Delta = \Delta_{\rm bare}$, derived from uncorrected spectral data. Figure~\ref{fig_11_convergence_FL} illustrates that the noiseless simulation results (blue open diamonds) consistently show a reduction in Trotter truncation error as the Trotter depth $M$ increases, indicating convergence. Notably, achieving convergence requires deeper circuits as the qubit number $N$ grows. This trend is consistent with the scaling of the Trotter depth cutoff $M_c$ [Eq.~\eqref{Trotter_depth_cutoff}], where $||[H_1,H_2]|| \leq 4|Jh|(N-1)$ for the TFIM~\cite{Lee2024}. In contrast, the noisy simulation results (green open squares) fail to converge, with errors increasing at higher $M$ values. As discussed in Sec.~\ref{sec_Markovian}, this behavior arises from the cumulative effect of Markovian noise channels when the noise probability exceeds the threshold. The real-device demonstrations (red open circles) exhibit even poorer convergence compared to the noisy simulations. Nevertheless, all outcomes for $\Delta = \Delta_{\rm corr}$ consistently validate the algorithm's robustness following baseline correction.

\section{Supplementary simulation data with $\mathcal{F} = \mathcal{F}_{\rm G}$}
\label{appendix_Gaussian_filter}

\begin{figure*}[t]
\begin{center}
\includegraphics[width=0.96\textwidth]{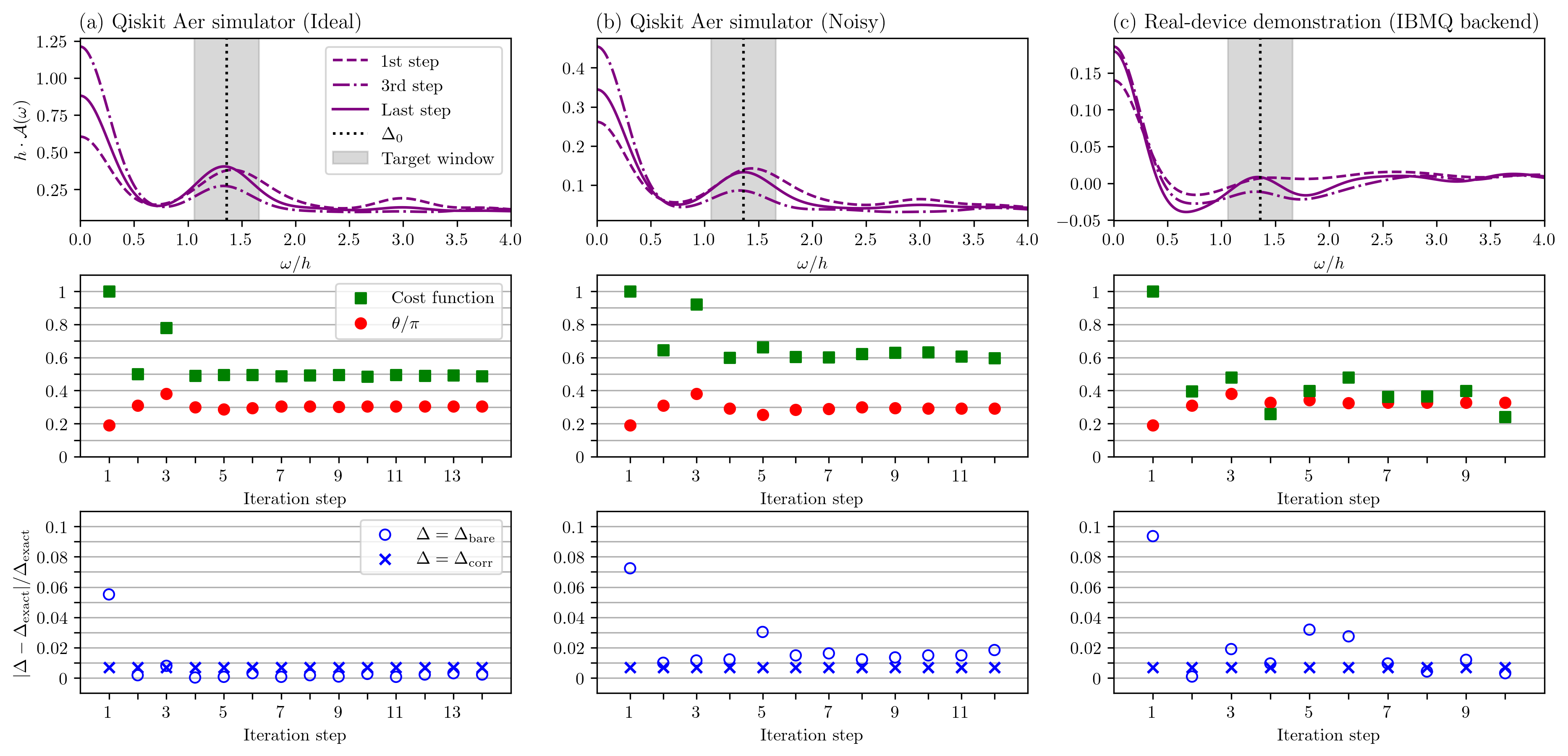}
\end{center}
\vspace{-15pt}
\caption{Simulation results demonstrating trial-state optimization in QGE for $\mathcal{F}=\mathcal{F}_{\rm G}$. Other settings are the same as in Fig.~\ref{fig_5_optimization_N5_FL}.}
\label{fig_12_optimization_N5_FG}
\end{figure*}

\begin{figure*}[t]
\begin{center}
\includegraphics[width=0.96\textwidth]{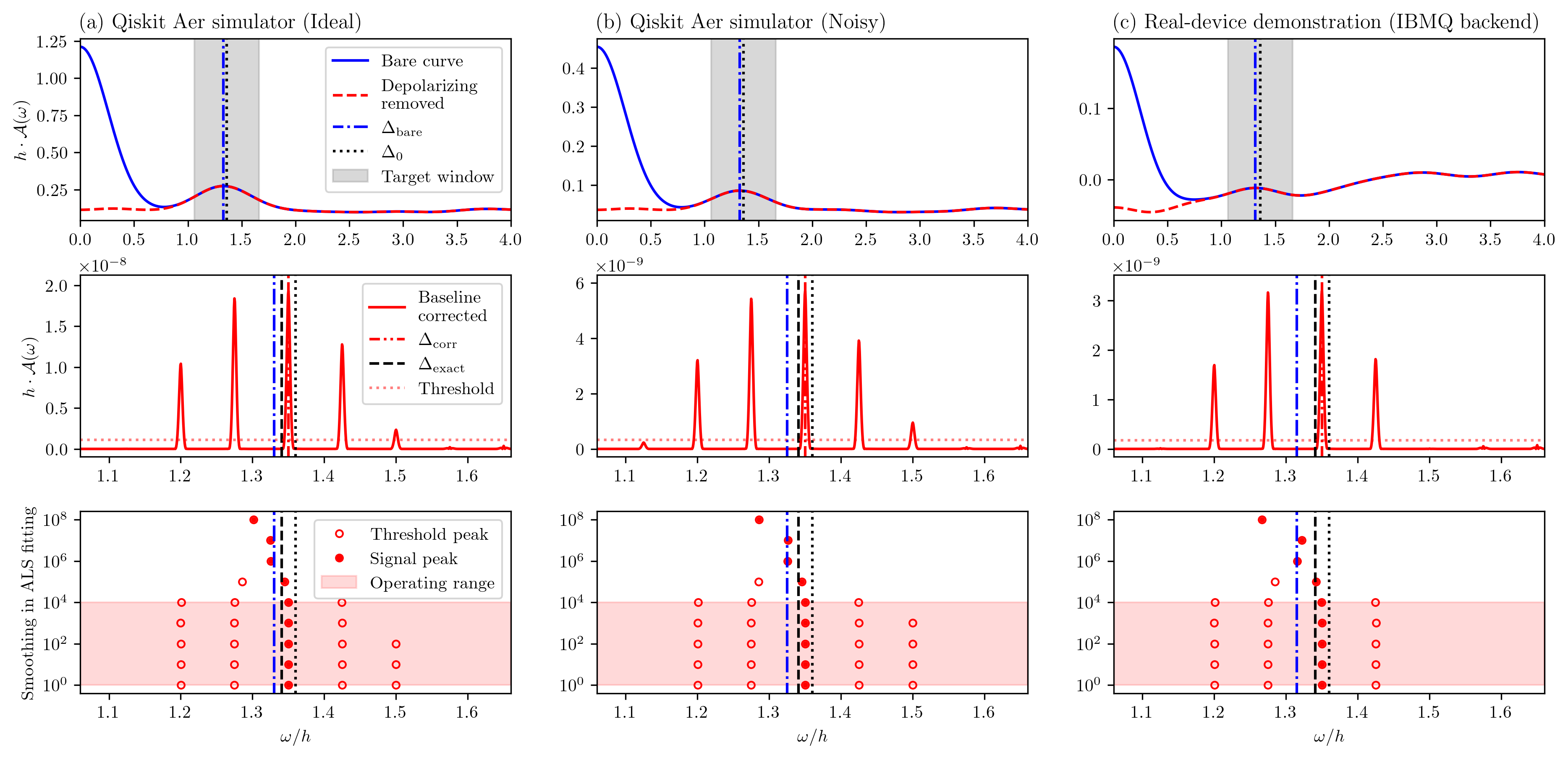}
\end{center}
\vspace{-15pt}
\caption{Simulation results illustrating the QGE procedure for $\mathcal{F}=\mathcal{F}_{\rm G}$. Other settings are the same as in Fig.~\ref{fig_10_ALSfitting_N5_FL}.}
\label{fig_13_ALSfitting_N5_FG}
\end{figure*}

\begin{figure*}[t]
\begin{center}
\includegraphics[width=0.92\textwidth]{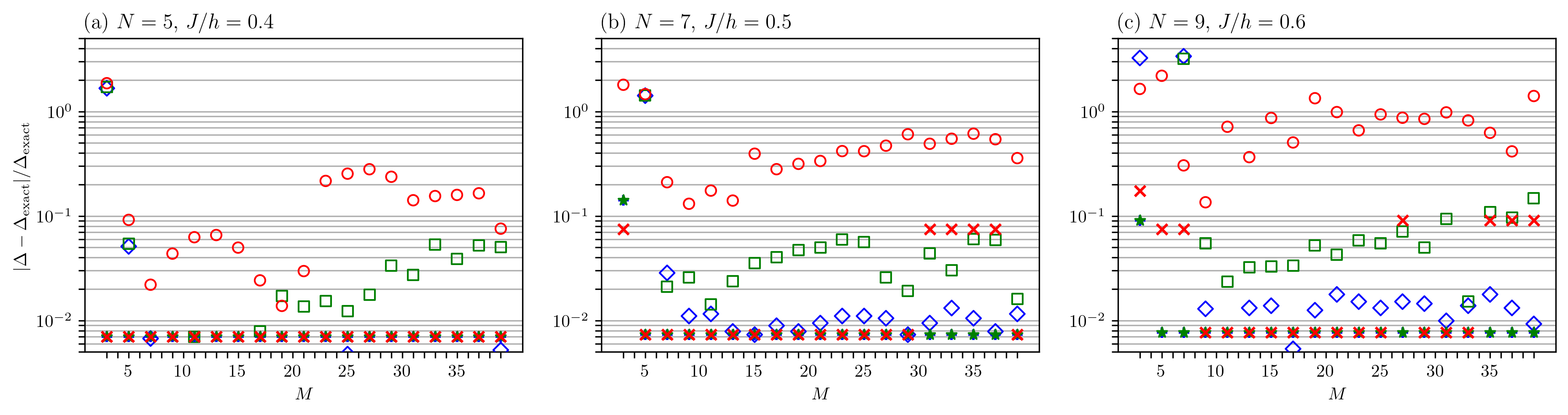}
\end{center}
\vspace{-15pt}
\caption{Simulation results for demonstrating the scalability and robustness of QGE for $\mathcal{F}=\mathcal{F}_{\rm G}$. Other settings are the same as in Fig.~\ref{fig_11_convergence_FL}.
}
\label{fig_14_convergence_FG}
\end{figure*}

In this appendix, we supplement the simulation data by applying the Gaussian filter to further validate the algorithm's performance. Figures~\ref{fig_12_optimization_N5_FG}-\ref{fig_14_convergence_FG} correspond to Figs.~\ref{fig_5_optimization_N5_FL}, \ref{fig_10_ALSfitting_N5_FL}, \ref{fig_11_convergence_FL}, respectively. While the overall features remain consistent across both filters, differences in finer details arise from the Gaussian line shape’s greater central concentration compared to the Lorentzian line shape. This characteristic suppresses interference from moderately or widely separated neighboring peaks.

\clearpage

\bibliography{refs}

\begin{thebibliography}{81}
\bibitem [{Com()}]{Comment_depolarizing}%
  \BibitemOpen
  \bibinfo {note} {For a quantum state $\rho$, the maximally mixed state is
  defined by the identity: $\frac{1}{d^2}\sum_{k=1}^{d^2}
  \big(\prod_{j=0}^{N-1} K_{k,j}\big)\rho\big(\prod_{j'=0}^{N-1} K_{k,j'}\big)
  = \frac{1}{d}I_d$, where $K_{k,j} \in\{I_2, \sigma^x, \sigma^y, \sigma^z\}$,
  $I_d = {\rm diag}(1,1,\cdots,1)$ of size $d = 2^N$, and
  $(\sigma^x,\sigma^y,\sigma^z)$ are Pauli matrices.}\BibitemShut {Stop}%
\end{thebibliography}

\end{document}